\documentclass[aps,prl,showpacs,twocolumn,superscriptaddress,floatfix]{revtex4-2}
\usepackage[utf8]{inputenc}
\newcommand{\figurescale}{1}
\usepackage{verbatim}
\usepackage{amsmath}
\usepackage{mathtools}

\DeclarePairedDelimiterX\braket[2]{\langle}{\rangle}{#1 \delimsize\vert #2}
\usepackage[margin=0.85in]{geometry} 

\usepackage[pdftex]{graphicx}
\usepackage[separate-uncertainty=true]{siunitx}
\DeclareSIUnit{\rpm}{rpm}
\usepackage[colorlinks=true,urlcolor=blue,linkcolor=blue,citecolor=blue]{hyperref}
\usepackage[usenames,dvipsnames]{xcolor}
\usepackage[T1]{fontenc}
\usepackage{placeins}
\usepackage{nicefrac}
\usepackage{braket}
\usepackage{pdfcomment}
\usepackage{xcolor}
\usepackage{braket}
\usepackage[normalem]{ulem}

\usepackage{cleveref}
\hypersetup{colorlinks,
	linkcolor={blue!75!black!80!yellow},
	citecolor={blue!75!black!80!yellow},
	urlcolor={blue!75!black!80!yellow}
}

\makeatletter
\renewcommand\@make@capt@title[2]{%
    \@ifx@empty\float@link{\@firstofone}{\expandafter\href\expandafter{\float@link}}%
    \sffamily{\textbf{#1}}\@caption@fignum@sep#2
}%
\renewcommand\figurename{Figure}
\makeatother

\usepackage{lineno}

\begin{document}

%############################## TITLE #########################################
\title{Atomistic spin textures on-demand in the van der Waals layered magnet CrSBr}
%##############################################################################
%
%############################ AUTHORS #########################################
\author{J.~Klein}\email{jpklein@mit.edu}
\affiliation{Department of Materials Science and Engineering, Massachusetts Institute of Technology, Cambridge, Massachusetts 02139, USA}
\author{T.~Pham}
\affiliation{Department of Materials Science and Engineering, Massachusetts Institute of Technology, Cambridge, Massachusetts 02139, USA}
\author{J.~D.~Thomsen}
\affiliation{Department of Materials Science and Engineering, Massachusetts Institute of Technology, Cambridge, Massachusetts 02139, USA}
\author{J.~B.~Curtis}%\email{jcurtis@seas.harvard.edu}
\affiliation{John A. Paulson School of Engineering and Applied Sciences, Harvard University, Cambridge, MA, USA}
\affiliation{Department of Physics, Harvard University, Cambridge, MA, USA}
\author{M.~Lorke}
\affiliation{Institut für Theoretische Physik, Universität Bremen, P.O. Box 330 440, 28334 Bremen, Germany}
\author{M.~Florian}
\affiliation{Institut für Theoretische Physik, Universität Bremen, P.O. Box 330 440, 28334 Bremen, Germany}
\author{A.~Steinhoff}
\affiliation{Institut für Theoretische Physik, Universität Bremen, P.O. Box 330 440, 28334 Bremen, Germany}
\author{R.~A.~Wiscons}
\affiliation{Department of Chemistry, Columbia University, New York 10027, United States}
\author{J.~Luxa}
\affiliation{Department of Inorganic Chemistry, University of Chemistry and Technology Prague, Technická 5, 166 28 Prague 6, Czech Republic}
\author{Z.~Sofer}
\affiliation{Department of Inorganic Chemistry, University of Chemistry and Technology Prague, Technická 5, 166 28 Prague 6, Czech Republic}
\author{F.~Jahnke}
\affiliation{Institut für Theoretische Physik, Universität Bremen, P.O. Box 330 440, 28334 Bremen, Germany}
\author{P.~Narang}\email{prineha@seas.harvard.edu}
\affiliation{John A. Paulson School of Engineering and Applied Sciences, Harvard University, Cambridge, MA, USA}
\author{F.~M.~Ross}\email{fmross@mit.edu}
\affiliation{Department of Materials Science and Engineering, Massachusetts Institute of Technology, Cambridge, Massachusetts 02139, USA}

%
%##############################################################################
%
\date{\today}
%
%##############################################################################
%									ABSTRACT
%##############################################################################
%
\begin{abstract}
Controlling magnetism in low dimensional materials is essential for designing devices that have feature sizes comparable to several critical length scales that exploit functional spin textures, allowing the realization of low-power spintronic and magneto-electric hardware.~\cite{uti.2004} Unlike conventional covalently-bonded bulk materials, van der Waals (vdW)-bonded layered magnets~\cite{huang_layer-dependent_2017,gong_discovery_2017,deng_gate-tunable_2018} offer exceptional degrees of freedom for engineering spin textures.~\cite{burch_magnetism_2018} However, their structural instability has hindered microscopic studies and manipulations. Here, we demonstrate nanoscale structural control in the layered magnet CrSBr creating novel spin textures down to the atomic scale. We show that it is possible to drive a local structural phase transformation using an electron beam that locally exchanges the bondings in different directions, effectively creating regions that have vertical vdW layers embedded within the horizontally vdW bonded exfoliated flakes. We calculate that the newly formed 2D structure is ferromagnetically ordered in-plane with an energy gap in the visible spectrum, and weak antiferromagnetism between the planes. Our study lays the groundwork for designing and studying novel spin textures and related quantum magnetic phases down to single-atom sensitivity, potentially to create on-demand spin Hamiltonians probing fundamental concepts in physics,~\cite{Ising.1925,Bethe.1931,Lieb.1961,mermin_absence_1966,Hohenberg.1967} and for realizing high-performance spintronic, magneto-electric and topological devices with nanometer feature sizes.~\cite{Basov.2017,Narang.2020}

\end{abstract}
%
%##############################################################################
%
\maketitle
%
%###############################################################################
%								MAIN TEXT
%###############################################################################
%

%\section{Introduction}

The observation of long-range magnetic order in the van der Waals (vdW) magnets CrI$_3$~\cite{huang_layer-dependent_2017}, Cr$_2$Ge$_2$Te$_6$~\cite{gong_discovery_2017} and Fe$_3$GeTe$_2$~\cite{deng_gate-tunable_2018} has expanded material platforms for the study of low-dimensional magnetism. Such pioneering work has strong relevance for exploring fundamental questions in physics, including probing the Mermin-Wagner-Hohenberg theorem~\cite{mermin_absence_1966,Hohenberg.1967} and realizing spintronic and magneto-electric devices~\cite{Song.2018,Kim.2018,Wang.2018,Klein.2018,Jiang.2019,Song.2019} that allow voltage-controlled magnetic and spin related properties.~\cite{deng_gate-tunable_2018,huang_electrical_2018,jiang_electric-field_2018,Jiang_doping_2018} Microscopically, such properties are governed by local coupling between spins that are strictly connected to the underlying crystal lattice.~\cite{Muhlbauer.2009,Yu.2010} Thus, creating new degrees of freedom requires tailoring local structural properties.~\cite{Romming.2013} Beyond the influence of the periodic crystal structure on the local magnetic coupling, chemical and structural defects as well as interfaces between domains, twins or confined geometries have been shown to produce exotic low-dimensional spin textures with novel chiralities,~\cite{Yu.2010b,Li.2017,Matsumoto.2016,Jin.2017} but controlled engineering of such defect topologies has proven to be challenging for further advancement in this field.~\cite{Duine.2018}

%###################### Figure 1 ################################################
\begin{figure*}
	\scalebox{\figurescale}{\includegraphics[width=1\linewidth]{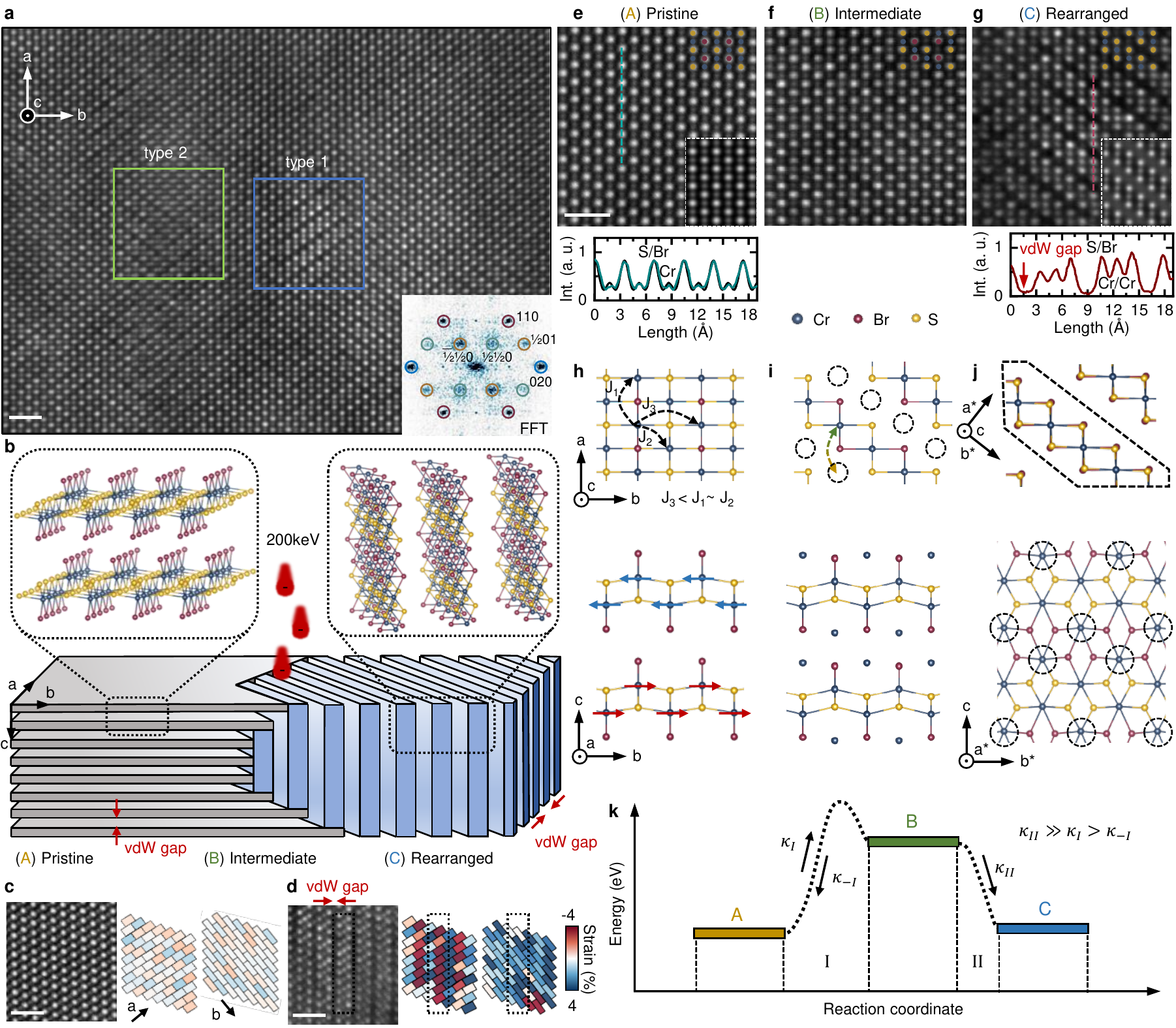}}
	\renewcommand{\figurename}{Fig.|}
	\caption{\label{fig1}
		\textbf{Electron beam-induced local crystal phase change in layered CrSBr.}
	    \textbf{a}. Low magnification STEM-HAADF image of a multilayer CrSBr flake imaged with an electron dose of $\sim\SI{1.3}{\per\pico\meter\squared\per\second}$ and a beam current of $\SI{60}{\pico\ampere}$ for $100s$ acquisition time (see Methods). Inset: Corresponding FFT showing double periodicity peaks (cyan and brown circles). The scale bar is $\SI{1}{\nano\meter}$.
	    \textbf{b}. Schematic illustration of the electron beam induced structural rearrangement of CrSBr from (A) pristine with out-of-plane stacking through (B) intermediate phase to (C) the rearranged structure exhibiting in-plane stacking.
		\textbf{c}. STEM-HAADF image of pristine multilayer CrSBr and corresponding strain maps along the $a$- and $b$-direction, obtained from S/Br atom column positions (see Methods).
		\textbf{d}. STEM-HAADF image of the rearranged crystal structure (C) and corresponding strain maps.	    
	    \textbf{e}. High magnification STEM-HAADF image of multilayer CrSBr. Bright and lighter spots correspond to S/Br and Cr atom columns, respectively. Experimental and simulated line profiles show the S/Br and the Cr columns. Inset: Multislice simulated STEM image of a pristine CrSBr sample (see Methods). Scale bar is $\SI{1}{\nano\meter}$.
	    \textbf{f}, STEM-HAADF image of multilayer CrSBr in the intermediate state (B).
	    \textbf{g}, A representative image of the rearranged CrSBr. Experimental and simulated line profiles show the migration of Cr. Inset: Corresponding multislice simulation STEM image of the rearranged structure. Scale bar is $\SI{1}{\nano\meter}$.
		\textbf{h}, Top-view of the crystal structure of CrSBr and exchange couplings J$_1$, J$_2$ and J$_3$ between next nearest neighbor Cr atoms and side-view of the bilayer with the AFM ordering.
		\textbf{i}, Top- and side-view of the intermediate crystal structure (B).
		\textbf{j}, Top- and side-view of the DFT relaxed crystal structure of rearranged CrSBr. 
		\textbf{k}, Reaction diagram of the electron beam induced structural rearrangement.
		}
\end{figure*}
%##############################################################################

In contrast to 3D materials, layered magnets offer superior design opportunities for engineering both periodic and non-periodic structures, and therefore spin textures, down to a single atomic level. Intrinsic (naturally occurring) spin textures below the critical temperature are made up of domain walls formed due to thermodynamically driven spin neutrality, while twin/grain boundaries and multicrystals are structural modulations that generally form naturally in structures to increase the entropy of the system. A hitherto overlooked aspect of vdW magnets is their potential for atomic scale structural modifications using electron or ion beam irradiation for the controlled creation of zero- to high dimensional defect topologies that range from single vacancies to locally induced crystal phase changes.~\cite{Lin.2014} Of particular interest is the exploration of interfaces and their proximity effects that provide a playground for studying local chirality,~\cite{Yu.2010b} magnetic anisotropy and spin canting on ultra-short length scales. Deterministic engineering of previously studied vdW magnets has remained elusive due to their poor structural stability, since full encapsulation and protection in a glove box environment and sophisticated fabrication methods are required.~\cite{deng_gate-tunable_2018,huang_layer-dependent_2017}

In this report, we demonstrate a versatile approach to engineer the structural landscape and the related spin textures in layered vdW magnets with single-atom precision. We conduct our study on the recently rediscovered vdW layered magnet CrSBr~\cite{Katscher.1966,Gser.1990,Wang.2019,Wang.2020} that constitutes an ideal candidate due to its known amenability for intrinsic structural changes~\cite{Gser.1990} and its highly promising electronic, optical and magnetic properties.~\cite{Katscher.1966,Gser.1990,Wang.2019,Wang.2020,Telford.2020,lee_magnetic_2020,Wilson.2021} CrSBr is an air-stable ferromagnetic (FM) insulator in the monolayer limit with a direct band gap of $\sim\SI{1.6}{\electronvolt}$~\cite{Telford.2020,Wang.2020,Wang.2019} hosting tightly bound magneto-excitons.~\cite{Wilson.2021} The magnetic easy axis is in-plane with an antiferromagnetic (AFM) interlayer coupling in the bulk with a N\'{e}el temperature of $T_N = \SI{132}{\kelvin}$~\cite{Gser.1990,lee_magnetic_2020} and a recently suggested intermediate soft magnetic phase up to $\SI{160}{\kelvin}$.~\cite{Telford.2020} The combination of these properties and the crystal stability makes CrSBr ideal for nanoscale structural modification, and hence the creation of spin textures with atomic resolution, a key requirement for building advanced spintronic and magneto-electric devices.

%###################### Figure 2 ################################################
\begin{figure*}
	\scalebox{\figurescale}{\includegraphics[width=0.75\linewidth]{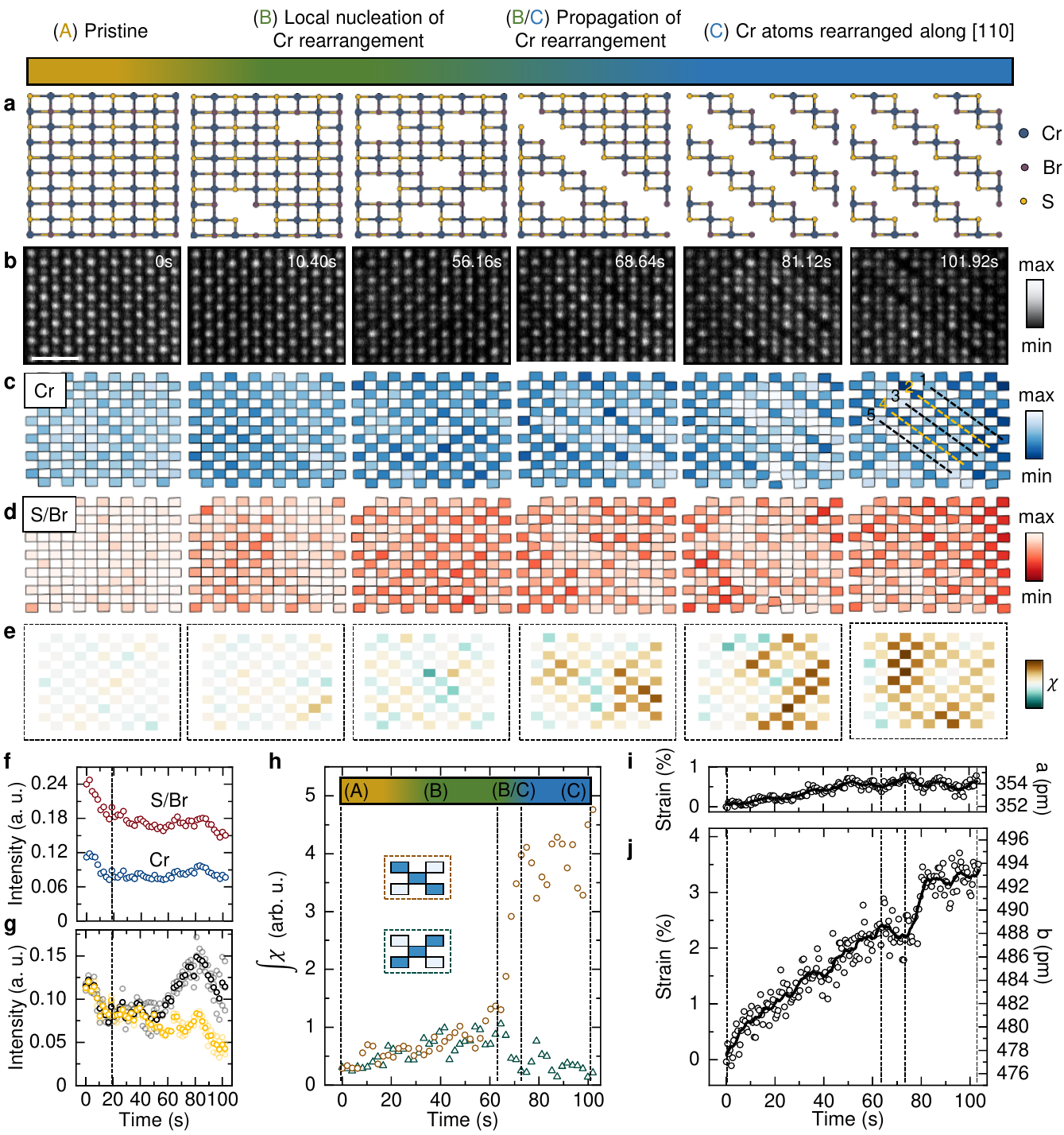}}
	\renewcommand{\figurename}{\textbf{Fig.|}}
	\caption{\label{fig2}
		\textbf{Nucleation and growth kinetics of new structural phases in CrSBr driven by electron beam irradiation.} 
		\textbf{a}, Schematic illustration (top-view) of the electron beam-induced structural rearrangement of Cr atoms.
		\textbf{b}, STEM-HAADF images of the CrSBr at $\SI{0}{\second}$, $\SI{10.40}{\second}$, $\SI{56.16}{\second}$, $\SI{68.64}{\second}$, $\SI{81.12}{\second}$ and $\SI{101.92}{\second}$. The flake imaged with an electron dose of $\sim\SI{1.6}{\per\pico\meter\squared\per\second}$ and a beam current of $\SI{75}{\pico\ampere}$. Scale bar is \SI{1}{\nano\meter}.
		Voronoi diagrams of \textbf{c} Cr and \textbf{d} S/Br from a Voronoi tesselation.
		\textbf{e}, Nucleation maps (degree of local nucleation, $\chi$) derived from the intensity difference of next nearest Voronoi cells along the two different Cr diagonals.
		\textbf{f}, Time evolution of the average column intensity of Cr (blue circles) and S/Br (red circles).
		\textbf{g}, Time evolution of averaged intensities from Cr diagonals on (black circles) and off (orange circles) the rearranged layer. See highlighted dashed lines in \textbf{c}.
		\textbf{h}, Time evolution of the integrated degree of nucleation. Inset: Schematic of the two different orientations along the diagonals. The pristine CrSBr (A) shows a local nucleation due reversible Cr migration (B) until $\sim\SI{62}{\second}$ where a sudden global ordering occurs (B/C) resulting in the rearrangement and relaxation into new 2D layers.
		\textbf{i}, \textbf{j} Time evolution of the net strain along the $a$- and the $b$-direction. The strain relaxation in the $b$-direction at $\sim\SI{62}{\second}$ is accompanied by the global crystal rearrangement. Scale bars are $\SI{1}{\nano\meter}$.
		}
\end{figure*}
%##############################################################################

\textbf{Local Crystal Phase Change.} We first demonstrate the electron beam-induced structural rearrangement. Fig.~\ref{fig1}a shows a scanning transmission electron microscopy high-angle annular dark-field (STEM-HAADF) image of a typical multilayer CrSBr crystal area after electron beam irradiation (See also Extended Data Figs. 1, 2 and 3). We observe two different types of new structures: type 1 (Supplementary Video 1) is the newly formed in-plane stacked 2D material, while type 2 (Supplementary Video 2) represents a layer stacking fault that forms due to a buildup of strain and potential local loss of Br (discussed in detail in the Extended Data Figs. 8, 9 and 15). We focus on structure type 1 throughout the remainder of this manuscript.

The local beam-induced structural rearrangement of the CrSBr crystal is illustrated in Fig.~\ref{fig1}b. Dosing pristine CrSBr (A) with $\SI{200}{\kilo\electronvolt}$ electrons results in the reversible migration of Cr atoms into the vdW gap to form a metastable intermediate crystal state (B) that eventually nucleates into a new magnetic (see below) 2D material (C) that is stacked along the in-plane direction. This striking rearrangement opens a vdW gap in the $a$-$b$-plane. While for pristine CrSBr we find lattice distances of $a = \SI{350.59}{\pico\meter}$ and $b = \SI{477.03}{\pico\meter}$, the vdW gap of the new structure is clearly visible from changes in lattice distances in the $a$- and $b$-direction as evidenced from strain maps of pristine (Fig.~\ref{fig1}c) and fully arranged CrSBr (Fig.~\ref{fig1}d) obtained from S/Br atom column positions. Generally, even pristine CrSBr crystals exhibit a large variation of up to $\pm 2\%$ over wide areas, likely due to the layered nature of the material (see Extended Data Fig. 5, 7 and 9 for detailed analysis of lattice distances and strain). A similar elongation along the $b$-direction is also observed in Li-intercalated FeOCl.~\cite{Palvadeau.1978,Beck.1990}

High-magnification STEM-HAADF images of the three different states (A, B and C) are shown in Fig.~\ref{fig1}e-g along with their corresponding top- and side-view crystal structure models in Fig.~\ref{fig1}h-j. To our advantage, in pristine CrSBr, columns containing S/Br and Cr atoms are spatially well separated in a top-view perspective along the $c$-direction and can be easily distinguished by their image contrast difference, in excellent agreement with the multislice simulations (see inset Fig.~\ref{fig1}e). The intensity variation of the atomic columns is also visible in the line profile in Fig.~\ref{fig1}g yielding a contrast of $\frac{I_{Cr}}{I_{S/Br}} \sim 80\%$. Irradiating with electrons creates a local and intermediate crystal phase (Fig.~\ref{fig1}f) visible from local intensity variations of the Cr atom column intensities which over a total of $\SI{80}{s}$ of exposure spontaneously rearranges into the new in-plane stacked 2D material (Fig.~\ref{fig1}g). 

The electron beam-driven mechanism of this structural rearrangement is described in a reaction coordinate diagram in Fig.~\ref{fig1}k. Local irradiation disrupts the bonding structure of Cr atoms, overcoming the activation energy such that the Cr can migrate into the vdW gap with a rate $\kappa_{I}$. This state is the local metastable intermediate state (B) (see Fig.~\ref{fig1}f and j). From this state, Cr can either reverse to its original position with a rate $\kappa_{-I}$ or irreversibly relax in the stable final rearranged state (C) at a rate $\kappa_{II}$. The top- and side-view of the newly formed 2D material is shown in Fig.~\ref{fig1}j. Cr atoms from every second diagonal migrate into a proximal diagonal. %In order to systematically describe the new crystal structure and related electronic and magnetic effects throughout this manuscript, we introduce a new coordinate system with $a = \hat{R} a^{*}$ and $b = \hat{R} b^{*}$ with $\hat{R} = \begin{pmatrix}
%\cos{\alpha} & -\sin{\alpha}\\
%\sin{\alpha} & \cos{\alpha} \end{pmatrix}$ being the rotational operator and $\alpha = \arctan{(\frac{b}{a}})$. 
From the side-view along the $a^{*}$ direction of a single new 2D layer (highlighted in Fig.~\ref{fig1}j), it is apparent that the migrated Cr atoms (see dashed circles) are bonded to four Br and two S atoms instead of the initial two Br atoms and four S atoms. The new crystal symmetry is similar to a 1\textit{T} phase of CrBr$_{2}$.~\cite{Kulish.2017} The experimental observation is in excellent agreement with the multislice simulated STEM image from the \textit{ab initio} relaxed structure model. The corresponding change in contrast can also directly be observed in the line cuts along the $a$-direction showing the net loss and surplus of Cr atoms in alternating Cr atom columns, clear evidence for the rearrangement and emergence of the new 2D material. Moreover, the change in the $a$- and $b$-directions is also in excellent agreement between experiment (Fig.~\ref{fig1}c and d) and our relaxed \textit{ab initio} DFT crystallographic model. We find that the ratio between experimental (theoretical) $a$- and $b$-directions increases from 1.3553 (1.3596) to 1.3946 (1.3834).

%###################### Figure 3 ################################################
\begin{figure*}
	\scalebox{\figurescale}{\includegraphics[width=0.75\linewidth]{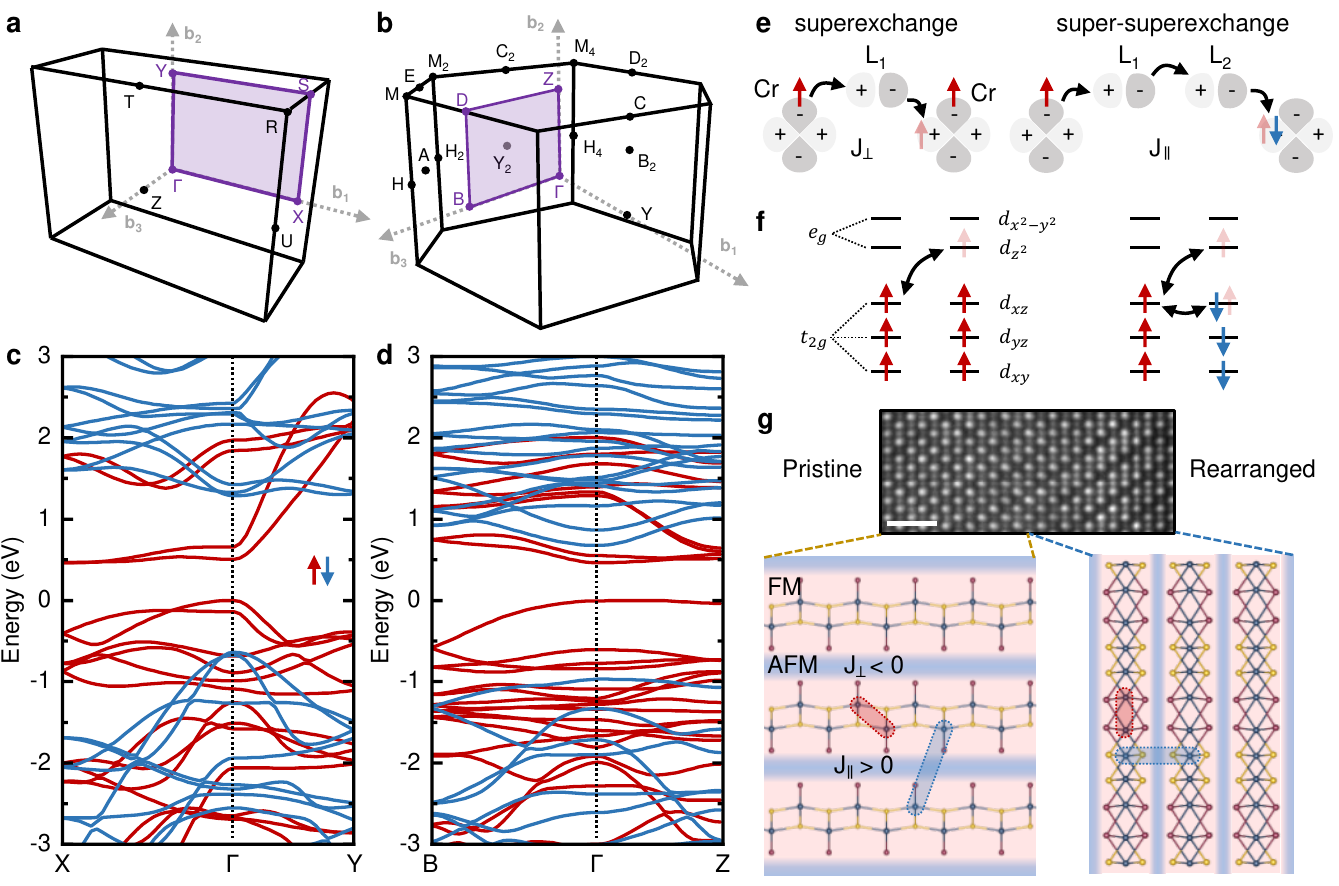}}
	\renewcommand{\figurename}{\textbf{Fig.|}}
	\caption{\label{fig3}
		\textbf{Electronic and magnetic properties of rearranged CrSBr.} 
		\textbf{a}, Brillouin zone of bulk \textbf{a} pristine and \textbf{b} rearranged CrSBr and their corresponding spin resolved DFT-PBE electronic band structure \textbf{c} and \textbf{d}.
		\textbf{e}, FM intralayer superexchange and AFM interlayer super-superexchange between the Cr atoms mediated by one or two ligands.
		\textbf{f}, Level scheme of intralayer superexchange interaction between Cr atoms. Crystal field splitting into doubly degenerate $e_g$ and triple degenerate $t_{2g}$ orbitals and the corresponding hopping terms for FM and AFM coupling.
		\textbf{g}, Pristine CrSBr exhibits intralayer FM order and interlayer AFM order for $T < T_N$. The rearranged CrSBr shows intralayer FM order and weak AFM order. Scale bar is \SI{1}{\nano\meter}.
		}
\end{figure*}
%##############################################################################

\textbf{Nucleation Kinetics.} We are particularly motivated by the possibility of forming these structural modifications within a matrix of the naturally grown structure because we can then utilize the spin structures of both structural domains to engineer low-dimensional chiral and/or topological spin textures at their interface. In order to analyze the time evolution of the process described in Fig.~\ref{fig1}k more quantitatively, we collect a sequence of $200$ STEM-HAADF images at a frame rate of $\SI{1.92}{}$ frames/s (total of $\SI{101.92}{\second}$) and an electron energy of $\SI{200}{\kilo\electronvolt}$ with a dose of $1.6 e^{-}\SI{}{\per\pico\meter\squared}$. The complete time evolution of the lattice rearrangement is shown in Fig.~\ref{fig2}a.

Selected STEM-HAADF images for $\SI{0}{\second}$, $\SI{10.40}{\second}$, $\SI{56.16}{\second}$, $\SI{68.64}{\second}$, $\SI{81.12}{\second}$ and $\SI{101.92}{\second}$ are shown in Fig.~\ref{fig2}b. Since the atom column intensity (image contrast) follows the dependence $I \propto Z \propto N$ ($Z$ is the atomic number of each element with the number of atoms $N$), we can track column intensities to directly infer changes in atom column stochiometry as a function of time. Column intensity maps obtained from a Voronoi tesselation of Cr and S/Br (Voronoi diagrams) are shown in Fig.~\ref{fig2}c and d (see Extended Data Figs. 4 and 10 for details on the analysis).

While both the Cr Voronoi diagram and the real space STEM-HAADF image reflect the time evolution of the rearranged structure, the S/Br atom columns do not exhibit such a redistribution of intensity over this time frame. Importantly, as the crystal undergoes rearrangement, there is no large change in overall stochiometry (see Fig.~\ref{fig2}f) but Cr atoms from diagonals (see dashed black lines 1, 3 and 5 in Fig.~\ref{fig2}c at $\SI{101.92}{\second}$) migrate to proximal diagonals (orange dashed lines 2 and 4). The migration becomes even more apparent by tracking the average intensity of the two different Cr diagonals as shown in Fig.~\ref{fig2}g. For $t > \SI{60}{\second}$, a splitting of the intensities occurs, a direct consequence of the electron beam-induced Cr migration. The striking migration of Cr in CrSBr is also further demonstrated by the filling of patterned 1D Cr vacancy line defects under irradiation (see Extended Data Fig. 16 and Supplementary Videos 3 and 4, respectively).

To study the nucleation kinetics and crystal rearrangement in space and time, we analyze Cr Voronoi cell intensities from Fig.~\ref{fig2}c using the next nearest neighbor Cr atom columns along the two different diagonals (see inset Fig.~\ref{fig2}h and Extended Data Fig. 11 for details). This analysis is sensitive to both local and global changes in the CrSBr nucleation kinetics. The spatial distribution of nucleation ($\chi$) at different times is shown in the corresponding nucleation maps (Fig.~\ref{fig2}e). Moreover, we derive and summarize the full time evolution from the integrated value of the nucleation maps (degree of nucleation) as shown in Fig.~\ref{fig2}h, which represents a precise measure on the degree of rearrangement along one diagonal or the other.

At $\SI{0}{\second}$ the CrSBr is pristine (A) with a homogeneous distribution of Cr atom column intensities. However, for $t > \SI{0}{\second}$ the degree of nucleation monotonically increases, reflecting local stochastic nucleation events of the Cr rearrangement (meta-stable state B) that is driven by the electron beam irradiation. Such events spatially appear as high values in the nucleation maps. For $t > \SI{60}{\second}$ a sudden global Cr rearrangement propagates through the analyzed area and the crystal fully rearranges into its final state (C). The full global rearrangement occurs on short time-scales ($\sim \SI{10}{\second}$). We observe the same trend in reciprocal space (see Extended Data Fig. 12). Moreover, aligning the scan-direction (e.g. along the [110] direction) manifests in an additional degree of freedom to control the preferred rearrangement direction (see Extended Data Fig. 3).

To further understand what microscopically triggers the sudden rearrangement, we quantify the time evolution of the net strain in the crystal along the $a$- and $b$-directions (see Fig.~\ref{fig2}i and further details on the analysis in the Extended Data Fig. 13). For the $a$-direction, the overall increase is less than $1\%$, which is due to the compensation of small and large strain values from the alternating lattice distortion within each newly formed 2D layer and the opening of the vdW gap (see Fig.~\ref{fig1}d). However, in the $b$-direction for $0 < t < \SI{60}{\second}$, we observe a continuous buildup of net strain that is accompanied by an increase in local nucleation (Cr migration events). A striking observation is that the net strain decreases simultaneously with the global rearrangement into the new 2D material. We interpret this as the lattice reaching a critical strain value due to the increased number of local Cr migration events that, in order to minimize its free energy, fully relaxes into the rearranged lattice structure (C) to release its accumulated strain.

%###################### Figure 4 ################################################
\begin{figure*}
	\scalebox{\figurescale}{\includegraphics[width=1\linewidth]{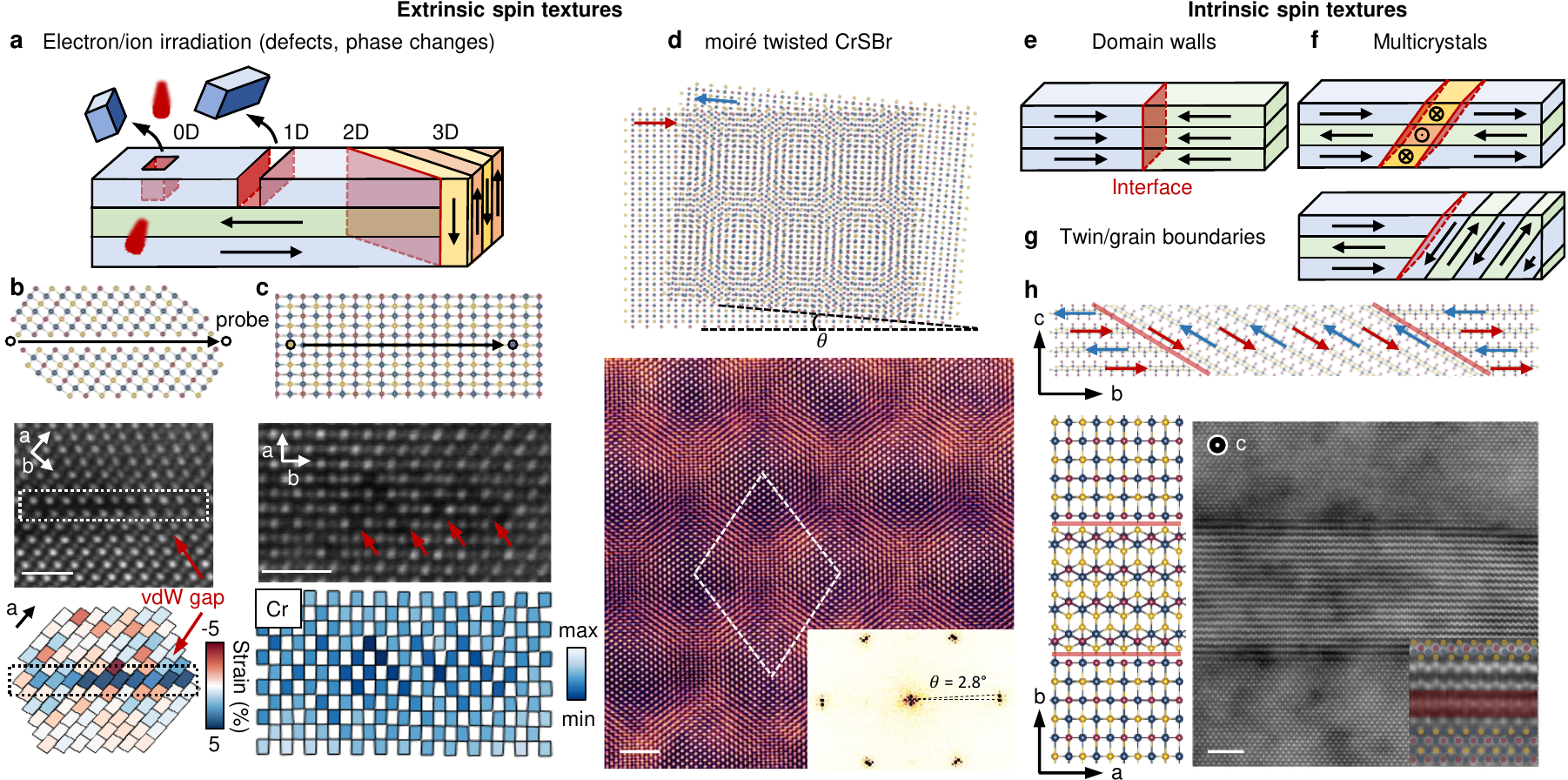}}
	\renewcommand{\figurename}{\textbf{Fig.|}}
	\caption{\label{fig4}
		\textbf{Spin textures on-demand.}
		\textbf{a}, Design of extrinsic spin textures by electron or ion irradiation via deterministic manipulation of crystal defects and phases. 
		\textbf{b}, A representative example showing the creation of a single Cr line vacancy by monitoring the scanning direction of the electron probe (along a pre-selected Cr line), introducing an artificial vdW gap as shown in the STEM-HAADF image and the corresponding local strain map. Scale bar is \SI{1}{\nano\meter}.
		\textbf{c}, Scanning along $b$-direction creates a zig-zag pattern due to selective migration of Cr atoms into the proximal vdW gap and local rearrangement. Scale bar is \SI{1}{\nano\meter}.
        \textbf{d} CrSBr moir\'e superlattice with controlled twist angles ($\sim2.8$° in this instance). Scale bar is \SI{2}{\nano\meter}.
		Intrinsic spin textures that naturally occur are \textbf{e}, domain walls, \textbf{f}, twin or grain boundaries and \textbf{g}, multicrystals. 
        \textbf{h}, Intrinsic spin texture induced by twinning in a four-layer CrSBr. The twin boundary is tilted by an angle of $\sim\SI{30.9}{\degree}$ with respect to the $c$-axis and creates an atomically defined interface. Scale bar is \SI{2}{\nano\meter}.
		}
\end{figure*}
%##############################################################################

\textbf{Electronic Structure and Magnetic Interactions.} To elucidate the electronic and magnetic properties of the rearranged structure, we calculate the electronic band structure and determine the magnetic interactions from a simplified tight-binding model of the relevant orbitals. Fig.~\ref{fig3}a and b show the full Brillouin zone of the pristine and rearranged CrSBr and their corresponding DFT calculated spin-polarized band structures for the essential high symmetry points are shown in Fig.~\ref{fig3}c and d. The lowest energy bands of the rearranged CrSBr are fully spin polarized with a direct gap transition at the $\Gamma$ point and an energy comparable to the pristine CrSBr suggesting a FM ground state. Since we are interested in the magnetic ordering of the rearranged CrSBr and in the spin texture the local rearrangement creates, we further determine the intralayer and interlayer exchange interaction for the pristine and the rearranged CrSBr. For our consideration we determine the intralayer exchange interaction $J_{\parallel}$ using superexchange mediated by the ligand and the interlayer interaction $J_{\perp}$ by super-superexchange via two ligands (see Fig.~\ref{fig3}e).
By performing strong-coupling perturbation theory (see Fig.~\ref{fig3}e,f) using the effective hoppings and crystal-field splittings extracted from spin-unpolarized {\it ab initio} band structure calculations, we indeed find the in-plane superexchange interaction is ferromagnetic, with a coordination and unit-cell averaged superexchange of order of $J_{\parallel} = -\SI{13}{\meV}$ for typical correlation parameters (here a minus sign indicates ferromagnetism).  

We also compute the interlayer magnetic interactions via the super-superexchange mechanism. Owing to the large separation between layers and roughly linear bond-angle, we find that the interlayer coupling is weakly antiferromagnetic, with a typical value being of order $J_{\perp} = \SI{1.5}{\meV}$. We therefore expect that for a relatively large range of temperatures the system exhibits quasi-two dimensional ferromagnetism, which is locally well-formed within the planes but only weakly coupled between the planes, creating unusual spin textures in interfaces between rearranged and pristine CrSBr (see Fig.~\ref{fig3}g). 

Note that, given the quasi-two dimensional nature of the rearranged structure, it is also important to determine the magnetocrystalline anisotropies, which arise from the spin-orbit corrections to superexchange. We can roughly estimate their size~\cite{Tartaglia.2020,Stavropoulos.2021} as follows.
Bromine has the largest atomic spin-orbit coupling, with $\xi_{\rm Br} \sim \SI{220}{\meV}$~\cite{Tartaglia.2020}.
This enters into the superexchange correction via second-order perturbation theory with a suppression by $\xi_{\rm Br}/\Delta_{\rm Br}\sim .06$, implying that the anisotropic exchange interactions should be of order $\xi_{\rm Br}/\Delta_{\rm Br} J_{\parallel} \sim .8\si{\meV}$, which is comparable to the interlayer exchange $J_{\perp}$.
We therefore expect that the true form of the magnetic ordering will be determined by a combination of the interlayer antiferromagnetism and in-plane anisotropies, and a more thorough analysis would be the subject of future calculations.

\textbf{Writing Spin Textures On-Demand.} The above calculations show that complex spin textures can be obtained by inserting, through targeted irradiation, regions of the new structure into a CrSBr matrix. The structural amenability of CrSBr provides a wide range of opportunities to program a variety of extrinsic spin textures into the lattice with a range of different topologies (0-3D) (see Fig.~\ref{fig4}a). To emphasize this point, we also demonstrate the flexibility of CrSBr for writing additional atom level structural modifications. Fig.~\ref{fig4}b shows a representative example of monitoring the scanning direction of the electron probe to selectively create a single Cr line vacancy, which artificially introduces a vertical vdW gap (see Fig.~\ref{fig4}b) within the horizontally vdW-bonded matrix. The absence of Cr atoms in the STEM-HAADF image and the formation of a vdW gap from the local strain analysis suggest the feasibility of modifying the crystal structure, and hence magnetic properties, on demand at the atomic scale. Scanning the electron probe across both S-Br and Cr atom columns creates a zig-zag pattern due to the selective migration of Cr atoms when the probe is on a Cr site. This is further shown by the variation in the analysed Cr atom column intensity. Another example for an extrinsic spin texture is the intentional introduction of a twist between two CrSBr sheets to create a superlattice (see Fig.~\ref{fig4}d). Twisted magnets are of particular interest for creating exotic quantum magnetic phases.~\cite{Hejazi.2020} Beyond the intentionally created textures are also intrinsic spin textures that can naturally occur and by themselves represent exciting magnetic systems given the layered nature of the host material. Examples include domain walls, multicrystals and twin and grain boundaries (see Fig.~\ref{fig4}e-g). Indeed, we observe examples of twin boundaries with a few $\SI{}{\nano\meter}$ width within exfoliated bulk crystals (see Fig.~\ref{fig4}h and Extended Fig. 14) creating atomically defined interfaces with different magnetic anisotropy on either side.

\textbf{Outlook.} We believe that these results lay the groundwork for creating and studying new structural phases and related spin textures in vdW materials on the atomic scale. Channel thicknesses of locally rearranged magnetic layers are shorter than state-of-the-art silicon transistor gate lengths, opening up exciting avenues for realizing new spintronic device designs combining out-of-plane and in-plane stacked 2D materials. Similarly, local writing of magnetic textures can be harnessed to controllably generate topological non-trivial spin textures with an unprecedented degree of control over dimensionality. Moreover, the intricate interplay of magnetization and localization effects of excitons will be interesting for future investigation, potentially leading to engineering spin-Hamiltonians \emph{in situ}.

The atomic level control demonstrated here will also be critical in a mechanistic understanding of proximity-induced effects in low dimensional heterostructures. Finally, magnetic and spin-orbit degrees of freedom in 2D magnets allow for flexible interaction engineering as well as simultaneous control of electronic band structure and topology, essential towards realizing universal classes of quantum many-body Hamiltonians and using vdW heterostructures as programmable quantum simulation platforms.

%
%##############################################################################
%               Acknowledgements & Contributions
%##############################################################################
%
\section{Acknowledgements}
J.K. acknowledges support by the Alexander von Humboldt foundation. T.P. and F.M.R. acknowledge the funding from the U.S. Department of Energy, Office of Basic Energy Sciences, Division of Materials Sciences and Engineering under Award DE‐SC0019336 for STEM characterization. J.D.T. acknowledges support from Independent Research Fund Denmark through grant no. 9035-00006B. Work by J.B.C. and P.N. is partially supported by the Quantum Science Center (QSC), a National Quantum Information Science Research Center of the U.S. Department of Energy (DOE). J.B.C. is an HQI Prize Postdoctoral Fellow and gratefully acknowledges support from the Harvard Quantum Initiative. P.N. is a Moore Inventor Fellow and gratefully acknowledges support through Grant GBMF8048 from the Gordon and Betty Moore Foundation. M.L, M.F., A.S. and F.J. were supported by the Deutsche Forschungsgemeinschaft (DFG) within RTG 2247 and through a grant for CPU time at the HLRN (Berlin/G\"ottingen). R.A.W. was supported by the Arnold O. Beckman Fellowship in Chemical Sciences. Z.S. and J.L. were supported by Czech Science Foundation (GACR No. 20-16124J).

\section{Author contributions}
J.K. and F.M.R. conceived and designed the experiments, J.D.T. and J.K. prepared the samples for electron microscopy, T.P. collected STEM data, J.K. wrote the scripts and analyzed the data, J.K., performed the optical measurements, R.A.W. contributed modelling the crystal structure, Z.S. and J.L. synthesized high-quality bulk CrSBr crystals, J.B.C. and P.N. computed the magnetic properties, M.L., M.F., A.S. and F.J. performed DFT calculations, J.K. wrote the manuscript with input from all co-authors. \\

%###############################################################################
%								Additional information
%###############################################################################
%

\section{Methods}

\subsection{Crystal synthesis}

CrSBr crystals were prepared by direct reaction from elements. Chromium (99.99\%, -60 mesh, Chemsavers, USA), bromine (99.9999\%, Sigma-Aldrich, Czech Republic) and sulfur (99.9999\%, Stanford Materials, USA) were mixed in stochiometric ratio in a quartz ampoule (35x220mm) corresponding to 15g of CrSBr. Bromine excess of 0.5g was used to enhance vapor transport. The material was pre-reacted in an ampoule using a crucible furnace at 700 °C for 12 hours, while the second end of the ampoule was kept below 250 °C. The heating procedure was repeated two times until the liquid bromine disappeared. The ampoule was placed in horizontal two zone furnace for crystal growth. First the growth zone was heated on 900 °C, while the source zone was heated on 700 °C for 25 hours. For the growth the thermal gradient was reversed and the source zone was heated from 900 °C to 940 °C and growth zone from 850 °C to 800°C over a period of 7 days. The crystals with dimensions up to 5x20 mm were removed from ampule in an Ar glovebox. 

\subsection{Structural characterization of bulk CrSBr}

We characterized grown CrSBr bulk crystals using SEM, EDS, XPS and XRD (see Extended Figs. 17, 18 and 19). The EDS data show homogeneous distribution of Cr, S and Br. High-resolution XPS reveals the presence of Br$^-$ and S$^{2-}$ anions and the Cr primarily in the $3+$ oxidation state. The X-ray diffraction reveals a pure single phase CrSBr with a high preferential orientation owing to the layered vdW structure.

\subsection{Sample fabrication}

Bulk CrSBr flakes are exfoliated using the Scotch tape method. Individual flakes and their thickness are determined by atomic force microscopy (AFM) and optical phase contrast. Exfoliated flakes were transferred to TEM compatible sample grids using cellulose acetate butyrate (CAB) as polymer handle.~\cite{Thomsen.2019} The CAB was then dissolved in acetone and the TEM grids rinsed in isopropanol before critical point drying. Twisted CrSBr samples are fabricated using the dry viscoelastic transfer method.

\subsection{Optical spectroscopy}

We optically characterized the CrSBr bulk crystal by Raman and photoluminescence (PL) spectroscopy at room temperature (see Extended Fig. 19). The PL spectrum of a multilayer CrSBr flake is measured exciting at $\SI{785}{\nano\meter}$ with an excitation power of $\sim \SI{50}{\micro\watt}$. For the Raman data, we used a low-frequency filter set to obtain both, Stokes and Anti-Stokes Raman modes. A monochromatic laser excitation at $\SI{532}{\nano\meter}$ was used at a power of $\SI{100}{\micro\watt}$.

\subsection{STEM imaging}

STEM imaging was performed with a probe-corrected Thermo Fisher Scientific Themis Z G3 $60$-$\SI{200}{\kilo\volt}$ S/TEM operated at $\SI{200}{\kilo\volt}$ and $\SI{60}{\kilo\volt}$ with the probe convergence semi-angle of 19 mrad and 30 mrad, respectively. The beam current was varied between $50$-$\SI{100}{\pico\ampere}$ to explore electron dosing conditions. A collection semi-angle of 63-200 mrad was used for HAADF STEM imaging. %EDS elemental maps were collected with a Thermo Fisher Scientific Super-X EDS detector and quantified using Velox software. 
%The flakes investigated had typical thicknesses of $5$ - $\SI{10}{\nano\meter}$. We observed the rearrangement phenomena down to four-layers in thickness.

\subsection{STEM data analysis}

In order to analyze sequences of STEM-HAADF images, movies are drift corrected before subsequent data processing. After drift correction, the position of the S/Br atom columns are determined by using atom-column indexing.~\cite{Sang.2014} Since Cr atom column intensities change throughout the video due to the Cr migration, we use the positions of proximal S/Br atom columns in order to approximate the Cr atom column position. Based on each set of x-y-coordinates from both, S/Br and Cr atom columns, we perform a Voronoi tesselation. Each Voronoi cell represents a single atom column. In order to obtain the atom column intensity, we integrate the intensity in the original STEM-HAADF image by masking out a polygon defined by each individual Voronoi cell. To determine local bond distances, we determine the distance between S/Br atom columns along the different crystallographic directions. Corresponding details are provided in Extended Figs. 4, 5, 6 and 7. 

\subsection{STEM multislice simulation}

STEM multislice simulations were performed on DFT relaxed crystal structure of the pristine and rearranged CrSBr using the Dr. Probe software. We used simulation parameters similar to the experimental conditions, including beam energy, convergence angle and collection angle, as well as the sample thickness. Simulated HAADF images were convolved with a Gaussian kernel with full-width at half-maximum of $\SI{70}{\pico\meter}$ to approximately account for the finite size of the effective electron source.

\subsection{\emph{Ab initio} calculations}
The ab initio calculations were performed using density functional theory (DFT), using the Vienna ab initio simulation package (VASP) and the projected augmented wave method \cite{Blochl:94,Kresse:99}. Bulk (vertical) structures were modelled with supercells containing 6 (12) atoms. 
The atomic and electronic structures were determined using the PBE functional. Van-der-Waals corrections were added in the Tkatchenko-Scheffler approximation \cite{Tkatchenko:09}. A plane wave basis with an energy cutoff of $Ecut = \SI{500}{\electronvolt}$ and a (7 × 5 × 2) ((2 × 4 × 4)) Monkhorst-Pack-point sampling has been employed for the bulk (vertical) structure. Structural relaxations were performed until the forces were smaller than $10^{-3}\SI{}{\electronvolt\per\angstrom}$. To generate the TB Hamiltonian, used in the spin-wave theory, the Wannier90 package~\cite{Wannier90} was employed.

\subsection{Magnetic Structure}

We predict the form of the magnetic structure by applying standard superexchange theory.
We model the onsite energy levels, including electron-electron interactions, by a Hubbard-Kanamori + crystal-field Hamiltonian~\cite{Georges.2013} which in the rotationally invariant limit assumes the form 
\begin{equation}
\label{eqn:SEJ}
    H = \sum_j  \sum_a \varepsilon_a^{(j)} \hat{n}_{ja} + \frac12 (U-\frac32 J_H )\hat{n}_j (\hat{n}_j -1) - J_H \hat{\mathbf{S}}_j^2 
\end{equation}
with Hubbard-$U$ modeling Coulomb repulsion, Hund's intra-atomic exchange $J_H$ which favors highest spin multiplicity, and single-particle crystal-field splitting $\varepsilon_a^{(j)}$.
Here we have introduced $d$-bands electronic operators $\hat{n}_{ja}\hat{n}_j,\hat{\mathbf{S}}_j$ which measure the total number of $d$-electrons in crystal-field level $\varepsilon_a^{(j)}$, the total $d$-electron occupation, and total spin of the Cr ion on site $j$, respectively.
The neglect of terms due to rotational symmetry breaking is a crude approximation, but partly justified by the small crystal-field splittings.
Following Ref.~\onlinecite{Wang.2019}, we compute the exchange for a range of values of $U = \SI{3.0}{\eV}-\SI{4.2}{\eV}$ and $J_H = \SI{.6}{\eV} - \SI{1.2}{\eV}$.
We obtain the crystal-field splittings from the {\it ab initio} calculations.

We determine the superexchange energies by performing strong-coupling perturbation theory up to second order in the effective $d-d$ hoppings.
See Ref.~\onlinecite{Besbes.2019} for a similar treatment in the case of CrI$_3$ and CrCl$_3$, and also Ref.~\onlinecite{Huang.2018}.
For a dimer of Cr ions in the $d^3d^3$ configuration, we find the superexchange interaction $J_{\rm se}$ by computing $J_{\rm se} = \frac29 (E_{FM} - E_{AFM})$, which assumes that the spin-interactions are a simple $SU(2)$ invariant (this is guaranteed in the absence of spin-orbit coupling) Heisenberg form, and FM and AFM indicate the states $|S = \frac32, m = \frac32\rangle \otimes |S = \frac32, m = \frac32\rangle $ and $|S = \frac32, m = \frac32\rangle \otimes |S = \frac32, m = -\frac32\rangle $, respectively.
In terms of the orbitally-resolved hoppings $t_{\beta\alpha}^{1 2}, t_{\beta\alpha}^{2 1}$ from orbital $\alpha$ on site 1 to orbital $\beta$ on site 2, and vice-versa, we find the result 
\begin{multline}
    J_{\rm se} = \frac23 \sum_{\alpha \beta}\left[ \frac13 \frac{ t_{\beta\alpha}^{1 2} t_{\beta \alpha}^{2 1} n_\alpha n_\beta}{U + 2J_H + \delta_{\beta \alpha}^{(21)} } -  \frac{J_H t_{\beta\alpha}^{1 2} t_{\beta \alpha}^{2 1} n_\alpha (1-n_\beta)}{(U  +\delta_{\beta \alpha}^{(21)} )^2 - (2J_H)^2} \right] \\
    + (1\leftrightarrow 2)
\end{multline}
where $\delta_{\beta \alpha}^{(21)} = \varepsilon_\beta^{(2)} - \varepsilon_\alpha^{(1)}$ is the energy of the $d-d$ excitation and $n_\alpha = \langle n_{j\alpha}\rangle$ is the occupation of the level $\alpha$ and is taken to be 1 for the lower three levels and 0 for the upper two levels.

The final ingredient is to derive the effective ligand-mediated hoppings which we approximate by 
\begin{equation}
    t^{12}_{\beta \alpha} = -\sum_{L} \left[ \hat{V}^{L\to 2}\cdot \hat{h}_L^{-1} \cdot \hat{V}^{1\to L}  \right]_{\beta\alpha},
\end{equation}
where the product is understood as a matrix product over the intermediate orbital states, and $L$ sums over the participating ligands (for edge sharing octahedra, there are two).
Here, $\hat{h}_L$ is the on-site ligand-$L$ block of the Wannier Hamiltonian, and $\hat{V}^{L\to 2}, \hat{V}^{1\to L}$ are the appropriate Cr-ligand hybridizations, also obtained from wannierized {\it ab initio} calculations.
We find that, except for extreme values of the Hubbard-$U$ and Hund's-$J_H$, the intra-layer exchanges are robustly FM.

In order to fix the interlayer ordering, we must compute the interlayer super-superexchange interactions.
In this case, we again apply Eqn.~\ref{eqn:SEJ}, but now using (alongside the direct $d$-band hoppings) the effective $d$-band hoppings of 
\begin{equation}
    t^{11}_{\beta \alpha} = + \sum_{L_1,L_2} \left[ \hat{V}^{2\leftarrow L_2}\hat{h}_{L_2}^{-1}\hat{V}^{L_2\leftarrow L_1}\cdot \hat{h}_{L_1}^{-1} \cdot \hat{V}^{L_1\leftarrow 1}  \right]_{\beta\alpha},
\end{equation}
with $L_1,L_2$ labeling all participating pairs of ligands. 
We find that this produces a weak AFM interlayer exchange $J_{\perp} \sim \SI{1.5}{\meV}$, as in the pristine case but now with a re-oriented magnetic structure. 

\section{Additional information}

\subsection{Supplementary Information} A detailed SI accompanies this paper along with electron microscopy videos.

\subsection{Data availability} The data that support the findings of this study are available from the corresponding author on reasonable request.

\subsection{Code availability} The codes used for data analysis as well as \emph{ab initio} calculations are available from the corresponding author on reasonable request.

\subsection{Competing financial interests} The authors declare no competing financial interests.

%
%###############################################################################
%								BIBLIOGRAPHY
%##############################################################################
%

\bibliographystyle{naturemag}
\bibliography{full}% Produces the bibliography via BibTeX.

\end{document}

%############################## TITLE #########################################
\title{Supplementary Information - Atomistic spin textures on-demand in the van der Waals layered magnet CrSBr}
%##############################################################################
%
%############################ AUTHORS #########################################
\author{J.~Klein}\email{jpklein@mit.edu}
\affiliation{Department of Materials Science and Engineering, Massachusetts Institute of Technology, Cambridge, Massachusetts 02139, USA}
\author{T.~Pham}
\affiliation{Department of Materials Science and Engineering, Massachusetts Institute of Technology, Cambridge, Massachusetts 02139, USA}
\author{J.~D.~Thomsen}
\affiliation{Department of Materials Science and Engineering, Massachusetts Institute of Technology, Cambridge, Massachusetts 02139, USA}
\author{J.~B.~Curtis}%\email{jcurtis@seas.harvard.edu}
\affiliation{John A. Paulson School of Engineering and Applied Sciences, Harvard University, Cambridge, MA, USA}
\affiliation{Department of Physics, Harvard University, Cambridge, MA, USA}
\author{M.~Lorke}
\affiliation{Institut für Theoretische Physik, Universität Bremen, P.O. Box 330 440, 28334 Bremen, Germany}
\author{M.~Florian}
\affiliation{Institut für Theoretische Physik, Universität Bremen, P.O. Box 330 440, 28334 Bremen, Germany}
\author{A.~Steinhoff}
\affiliation{Institut für Theoretische Physik, Universität Bremen, P.O. Box 330 440, 28334 Bremen, Germany}
\author{R.~A.~Wiscons}
\affiliation{Department of Chemistry, Columbia University, New York 10027, United States}
\author{J.~Luxa}
\affiliation{Department of Inorganic Chemistry, University of Chemistry and Technology Prague, Technická 5, 166 28 Prague 6, Czech Republic}
\author{Z.~Sofer}
\affiliation{Department of Inorganic Chemistry, University of Chemistry and Technology Prague, Technická 5, 166 28 Prague 6, Czech Republic}
\author{F.~Jahnke}
\affiliation{Institut für Theoretische Physik, Universität Bremen, P.O. Box 330 440, 28334 Bremen, Germany}
\author{P.~Narang}\email{prineha@seas.harvard.edu}
\affiliation{John A. Paulson School of Engineering and Applied Sciences, Harvard University, Cambridge, MA, USA}
\author{F.~M.~Ross}\email{fmross@mit.edu}
\affiliation{Department of Materials Science and Engineering, Massachusetts Institute of Technology, Cambridge, Massachusetts 02139, USA}
%
%##############################################################################
%
\date{\today}
%
%##############################################################################

%##############################################################################
%
\maketitle
%
%###############################################################################
%								MAIN TEXT
%###############################################################################
%

\tableofcontents

\newpage

%\sectionfont{\titlecap}

\section{Merged crystal structure.}

%
%###################### Figure Crystal structure ################################################
\begin{figure*}[ht]
\scalebox{\figurescale}{\includegraphics[width=1\linewidth]{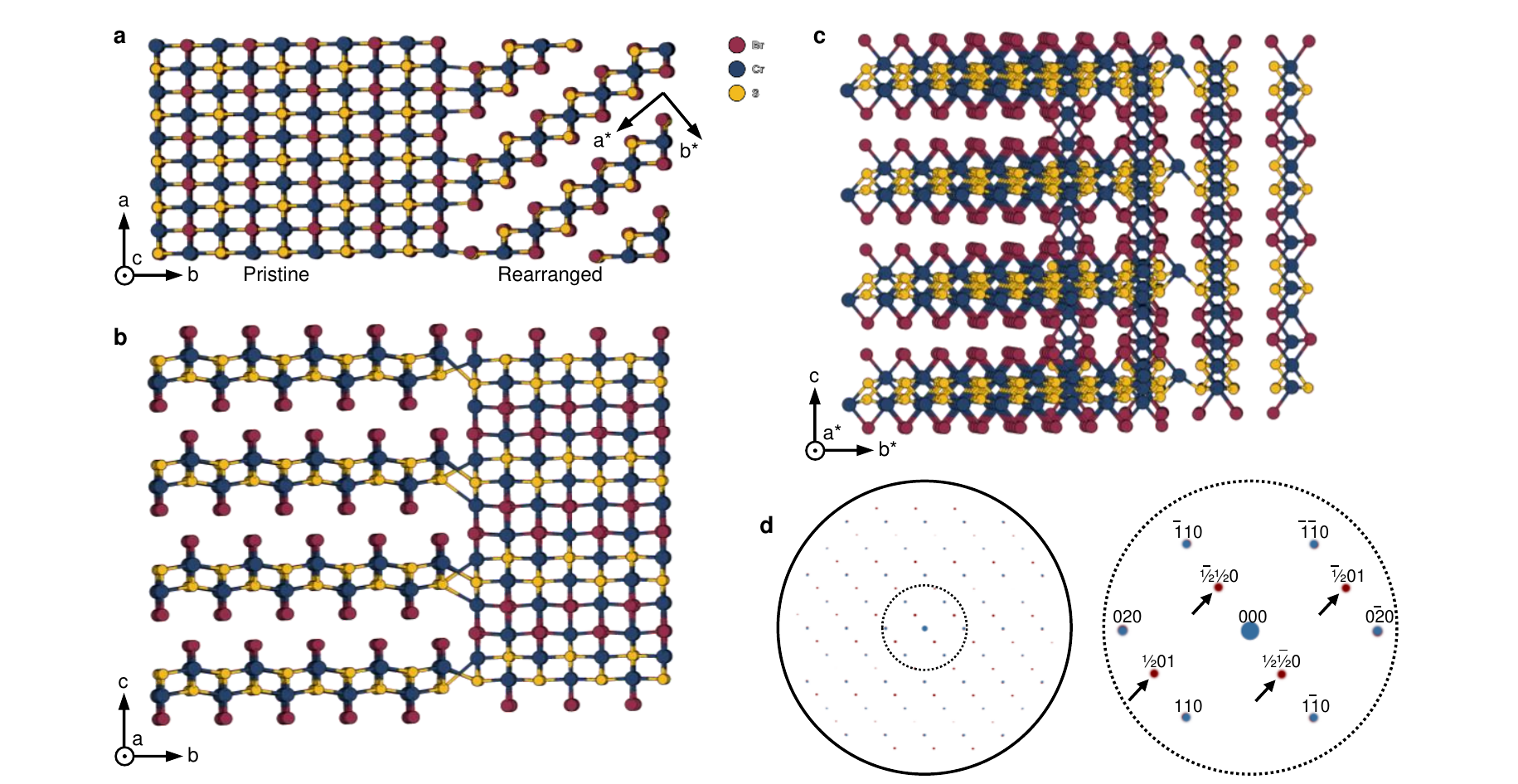}}
\renewcommand{\figurename}{Extended Data Fig.|}
\caption{\label{SIfigCrystal_Structure}
\textbf{Schematic diagrams of structures containing both pristine and rearranged CrSBr regions.} 
\textbf{a}, Top-view.
\textbf{b}, Side-view.
\textbf{c}, View-along the rearranged layers.
\textbf{d}, Simulated diffraction pattern. The arrowed spots originate from the rearranged structure.
}
\end{figure*}
%##############################################################################
%
\newpage

\section{Electron beam-induced structural changes.}

%
%###################### Figure Defect images low mag ################################################
\begin{figure*}[ht]
\scalebox{\figurescale}{\includegraphics[width=1\linewidth]{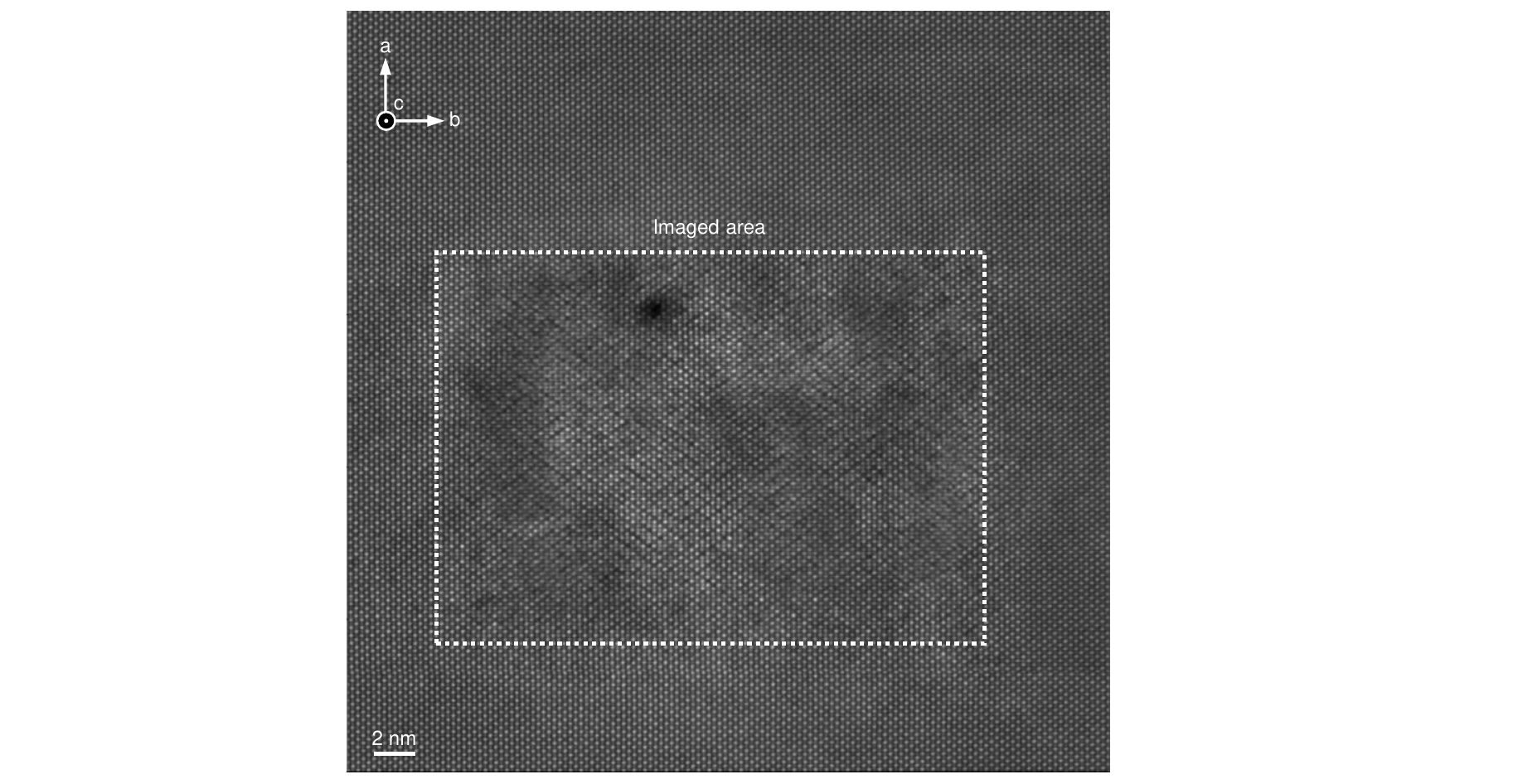}}
\renewcommand{\figurename}{Extended Data Fig.|}
\caption{\label{SIfigdefectimages}
\textbf{Electron beam-induced structural changes.} 
Low-magnification STEM-HAADF image at $\SI{200}{\kilo\electronvolt}$ of an electron beam dosed area.% with type 1 and type 2 defective structures.
}
\end{figure*}
%##############################################################################
%

\newpage

%
%###################### Figure Larger area ################################################
\begin{figure*}[ht]
\scalebox{\figurescale}{\includegraphics[width=1\linewidth]{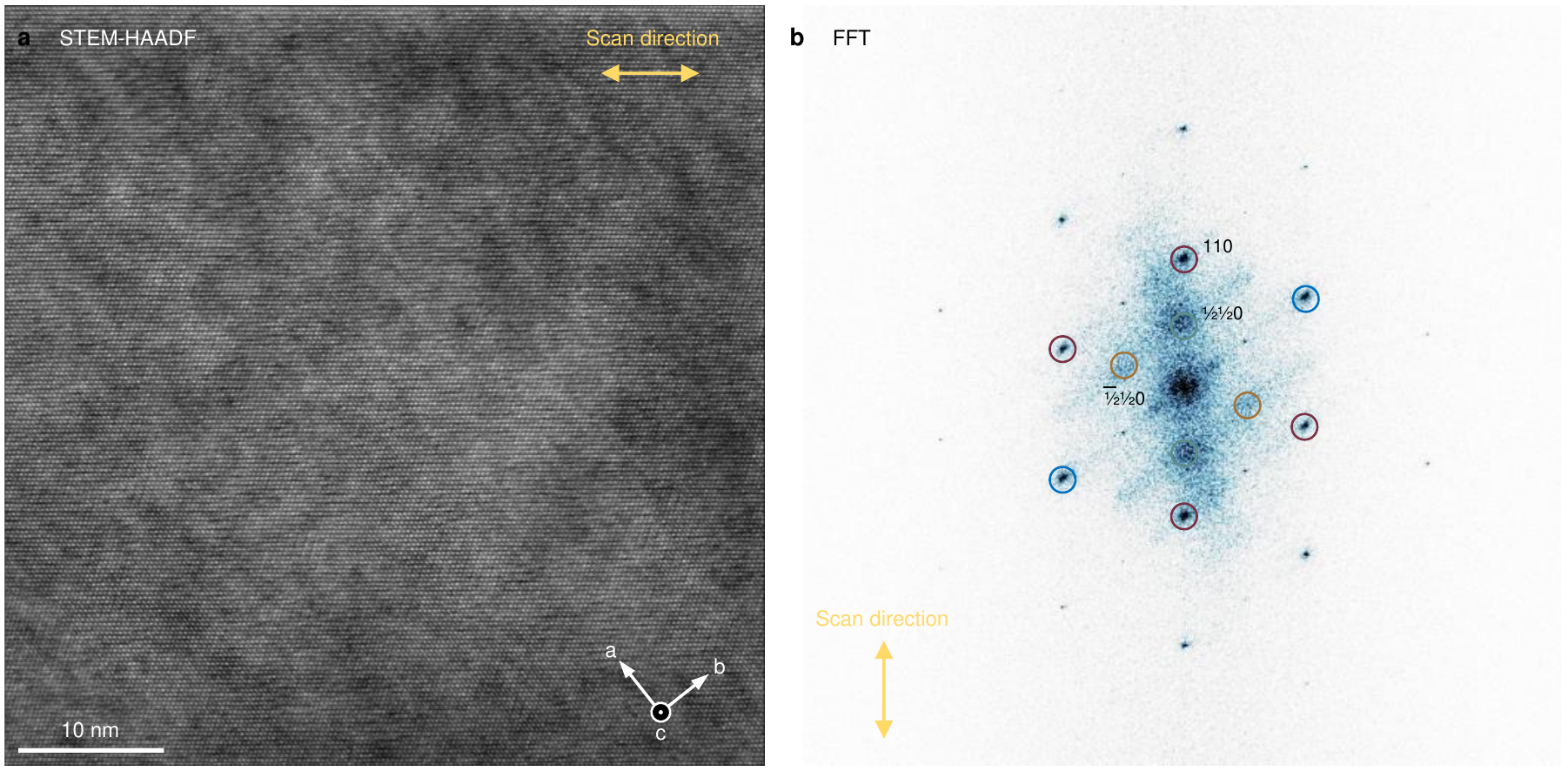}}
\renewcommand{\figurename}{Extended Data Fig.|}
\caption{\label{SIfiglargearea}
\textbf{Electron beam-induced structural changes in a large area.} 
\textbf{a}, Low-magnification STEM-HAADF image at $\SI{200}{\kilo\electronvolt}$ of an electron beam dosed area.
\textbf{b}, Corresponding FFT showing preferential rearrangement along the electron probe scan direction reflected by the higher intensity of the FFT peaks along the [110] direction. 
}
\end{figure*}
%##############################################################################
%

\newpage

\section{Voronoi cell intensity and strain image analysis.}

%
%###################### Figure 1 ################################################
\begin{figure*}[ht]
\scalebox{\figurescale}{\includegraphics[width=1\linewidth]{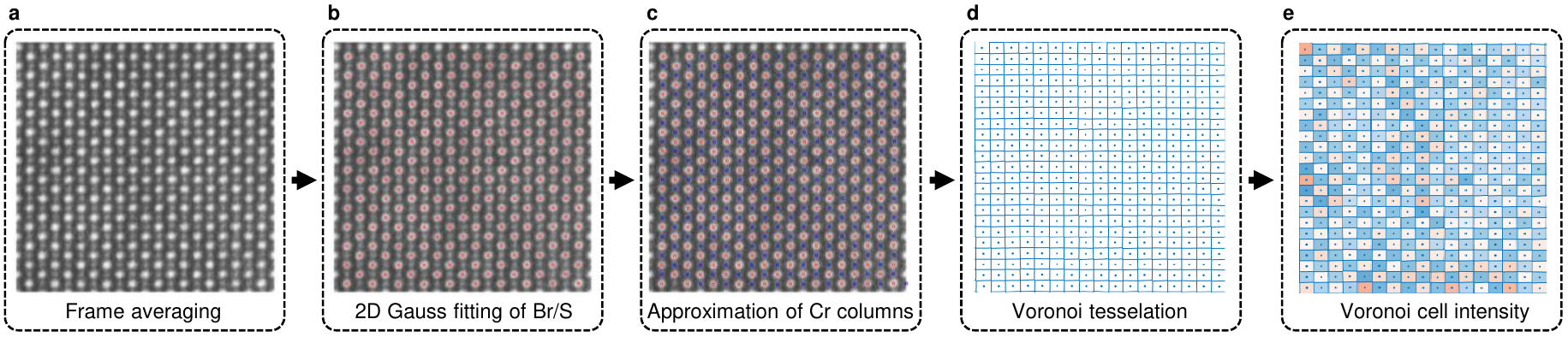}}
\renewcommand{\figurename}{Extended Data Fig.|}
\caption{\label{SIfig1}
\textbf{Quantitative STEM-HAADF image analysis using Matlab.} 
\textbf{a}, STEM-HAADF image of a $\sim \SI{10}{\nano\meter}$ thin CrSBr flake after drift correction and frame averaging. In this example, four frames were averaged.
\textbf{b}, Selective 2D Gauss fitting and indexing of the S/Br atom columns (red circles) using the atom column indexer.~\cite{Sang.2014}
\textbf{c}, Approximation of Cr atom column positions (blue circles) using four proximal S/Br positions that are obtained from the initial indexing through the 2D Gaussian fitting.
\textbf{d}, Voronoi diagram of both the Cr and S/Br atomic columns obtained from a Voronoi tesselation.
\textbf{e}, Corresponding Voronoi cell intensities of the S/Br (red) and Cr (blue) columns. The intensity of each Voronoi cell is obtained by masking out the polygon area from the initial image and by integrating the intensity of all pixels within the masked cell. 
}
\end{figure*}
%##############################################################################
%
\newpage
%
%###################### Figure 3 ################################################
\begin{figure*}[ht]
\scalebox{\figurescale}{\includegraphics[width=1\linewidth]{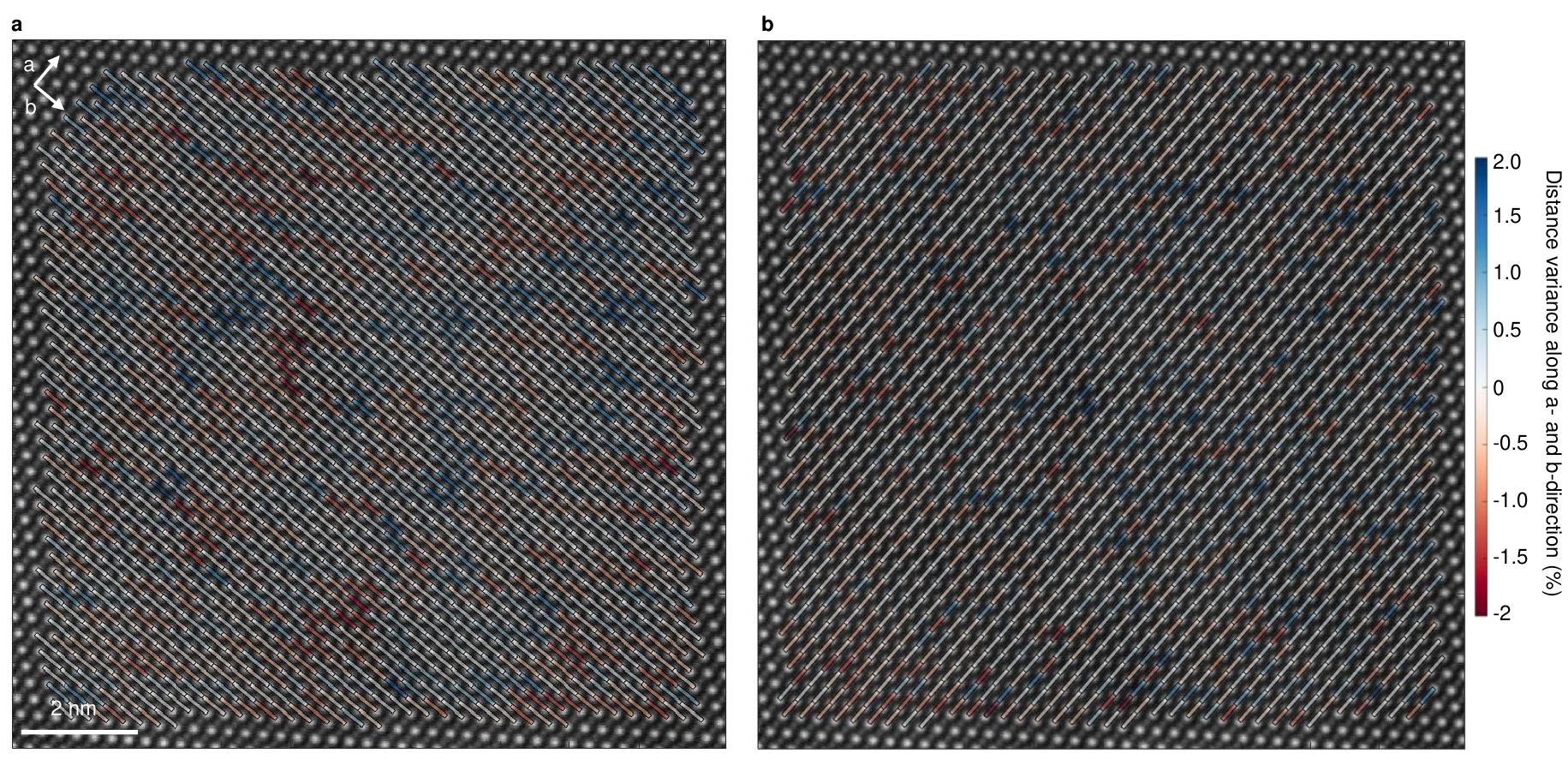}}
\renewcommand{\figurename}{Extended Data Fig.|}
\caption{\label{SIfig3}
\textbf{Local strain in CrSBr.} 
\textbf{a}, STEM-HAADF image of a pristine $\sim \SI{10}{\nano\meter}$ thin CrSBr flake prior to electron beam dosing.
\textbf{b}, \textbf{c}, Corresponding analysis of lattice distances along the a- and b-directions respectively, obtained from the indexed S/Br atom columns. The color of the bars plotted between the S/Br atom columns represent the deviation of the lattice parameter obtained from a normal distribution. The mean lattice parameters for the pristine image are $b = (477.82\pm3.03)\SI{}{\pico\meter}$ and $a = (346.46\pm2.14)\SI{}{\pico\meter}$ (see Extended Data Fig.~\ref{SIfig5})
}
\end{figure*}
%##############################################################################
%

\newpage

%
%###################### Figure 2 ################################################
\begin{figure*}[ht]
\scalebox{\figurescale}{\includegraphics[width=1\linewidth]{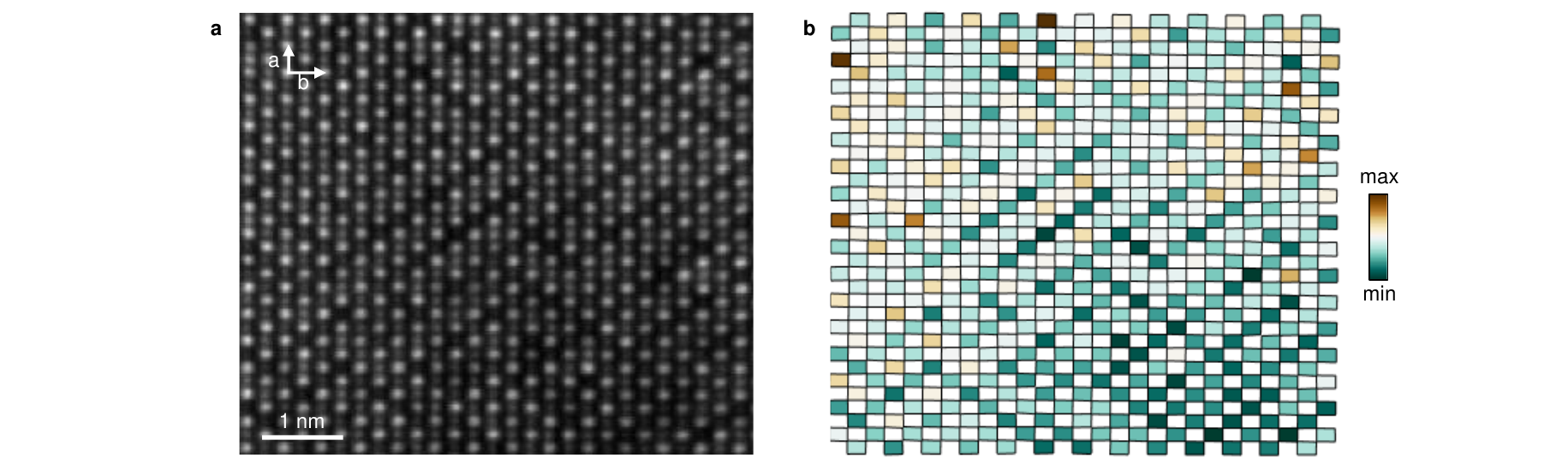}}
\renewcommand{\figurename}{Extended Data Fig.|}
\caption{\label{SIfig2}
\textbf{Visualizing Cr rearrangement via Voronoi analysis.} 
\textbf{a} STEM-HAADF image showing a maze-like pattern due to the rearrangement of Cr atoms between the columns along the diagonals ([110]) direction.
\textbf{b} Voronoi diagram of the corresponding intensity of the Cr columns as obtained from the quantitative image analysis described in Extended Data Fig.~\ref{SIfig1}.
}
\end{figure*}
%##############################################################################
%
\newpage

%
%###################### Figure 4 ################################################
\begin{figure*}[ht]
\scalebox{\figurescale}{\includegraphics[width=1\linewidth]{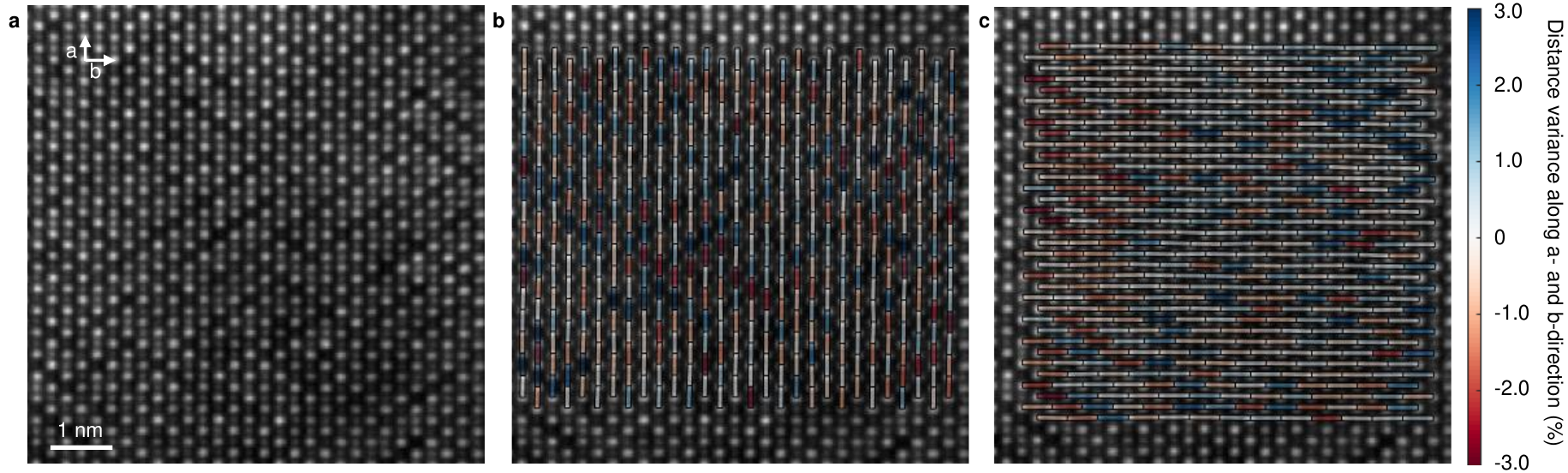}}
\renewcommand{\figurename}{Extended Data Fig.|}
\caption{\label{SIfig4}
\textbf{Local strain of the rearranged CrSBr.} 
\textbf{a}, STEM-HAADF image of an irradiated area revealing the maze-like structure of diagonal lines caused by the rearrangement of Cr atoms.
\textbf{b}, \textbf{c}, Corresponding analysis of lattice distances along the a- and b-direction, respectively, obtained from the indexed S/Br atom columns. The mean lattice parameters for the pristine image are $b = (495.99\pm5.82)\SI{}{\pico\meter}$ and $a = (348.90\pm4.98)\SI{}{\pico\meter}$ (see Extended Data Fig.~\ref{SIfig1})
}
\end{figure*}
%##############################################################################
%
\newpage
%
%###################### Figure strain ################################################
\begin{figure*}[ht]
\scalebox{\figurescale}{\includegraphics[width=1\linewidth]{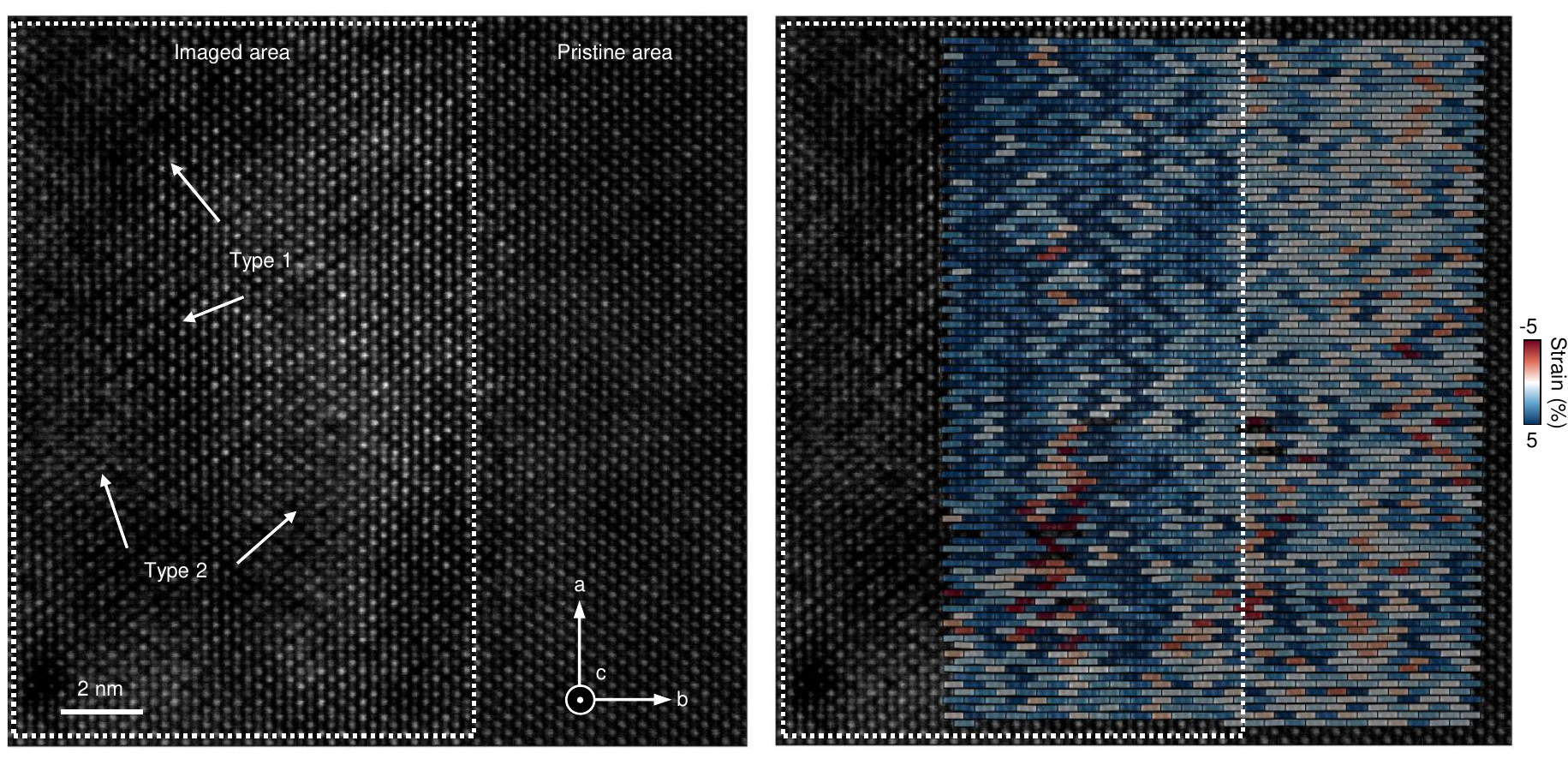}}
\renewcommand{\figurename}{Extended Data Fig.|}
\caption{\label{SIfigstrain}
\textbf{Local strain analysis of an interface between electron-dosed and pristine CrSBr.} 
Low-magnification STEM-HAADF image of an interface region between irradiated and pristine CrSBr and the corresponding atomistic strain analysis along the $b$-direction. The irradiated area shows type 1 and type 2 defects and increased strain along the $b$ direction. The stacking fault defect (type) likely occurs due to reduce strain as suggested by negative relative strain values in those regions.
}
\end{figure*}
%##############################################################################
%
\newpage
%
%###################### Figure 5 ################################################
\begin{figure*}[ht]
\scalebox{\figurescale}{\includegraphics[width=0.8\linewidth]{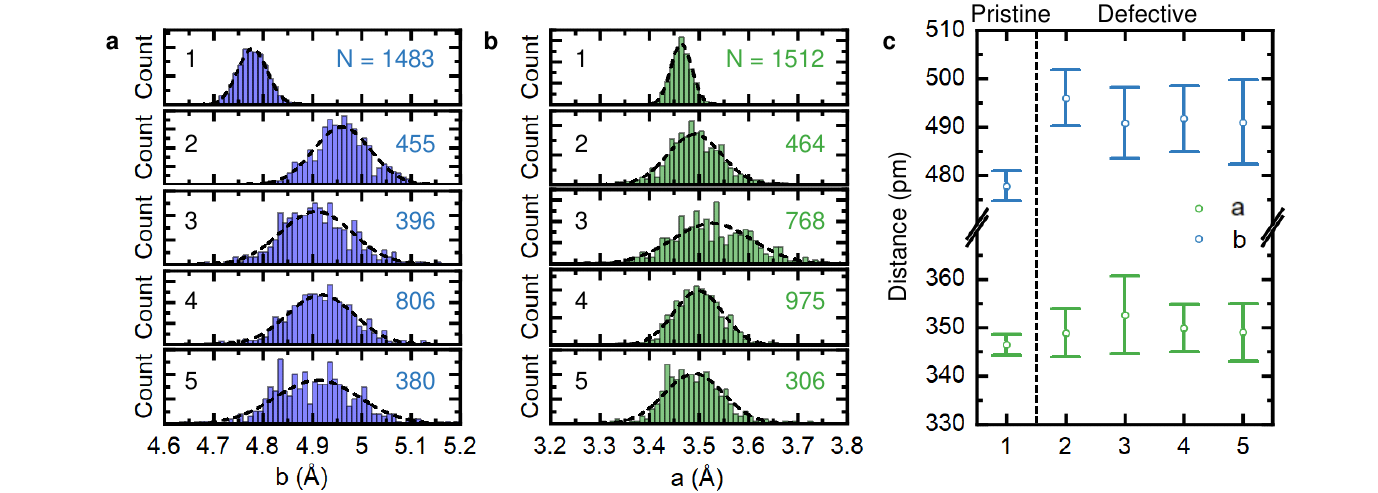}}
\renewcommand{\figurename}{Extended Data Fig.|}
\caption{\label{SIfig5}
\textbf{Statistics of lattice distances of pristine and defective CrSBr}
\textbf{a}, Lattice parameters along the a-direction between the S/Br columns from quantitative STEM analysis. The histogram number 1 is from a pristine area (see Extended Data Fig.~\ref{SIfig3}), while histograms number 2 - 5 are from several different areas after irradiation that reveal the 1D maze structure (see Extended Data Fig.~\ref{SIfig4}) The number in the top right reflects the ensemble size (number of S/Br - S/Br distances). The dashed black line is a fitted normal distribution. 
\textbf{b}, Lattice parameters along the $a$-direction between the S/Br columns from quantitative STEM analysis.
\textbf{c}, Comparison of $a$- and $b$-lattice distances obtained from fitted normal distributions from the histograms number 1 - 5 in \textbf{a} and \textbf{b}.
\textbf{d}, The data suggest an elongation along the a-direction by about $3\%$ while the b-direction stays constant within the statistical error.
}
\end{figure*}
%##############################################################################
%

\newpage

\section{Time evolution of nucleation kinetics.}

%
%###################### Figure 6 ################################################
\begin{figure*}[ht]
\scalebox{\figurescale}{\includegraphics[width=1\linewidth]{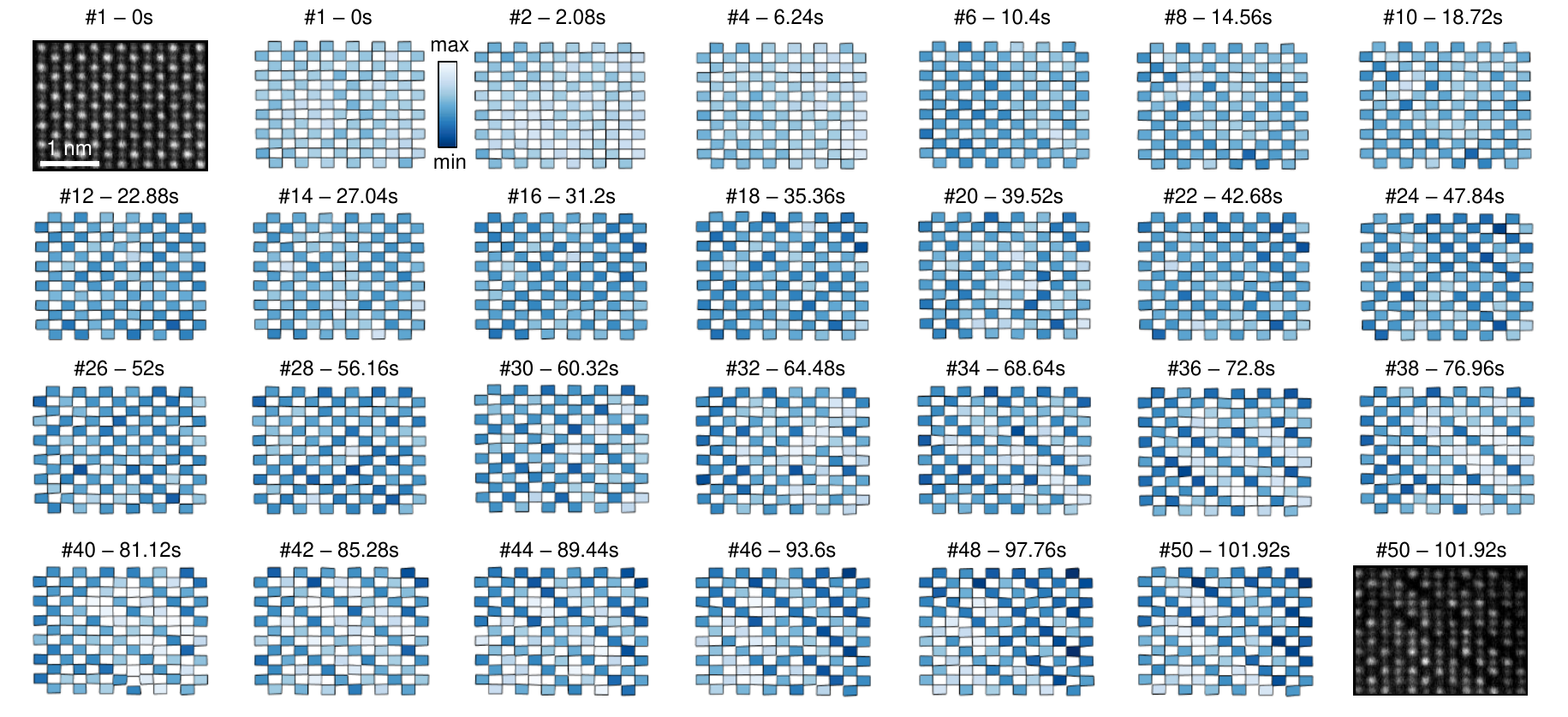}}
\renewcommand{\figurename}{Extended Data Fig.|}
\caption{\label{SIfig6}
\textbf{Time dependent kinetics of the Cr rearrangement obtained from a Voronoi analysis of STEM-HAADF images.}
Voronoi diagrams of the Cr atom columns for selected frames from a sequence of STEM-HAADF images taken at a frame rate of $\sim\SI{2}{\hertz}$. In order to increase the signal-to-noise ratio for the data analysis, four frames were averaged resulting in a frame rate of $\sim\SI{0.5}{\hertz}$ for the analysis. The analysis described in Extended Data Fig.~\ref{SIfig1} was applied to each averaged frame.
}
\end{figure*}
%##############################################################################
%
\newpage
%
%###################### Figure correlation ################################################
\begin{figure*}[ht]
\scalebox{\figurescale}{\includegraphics[width=1\linewidth]{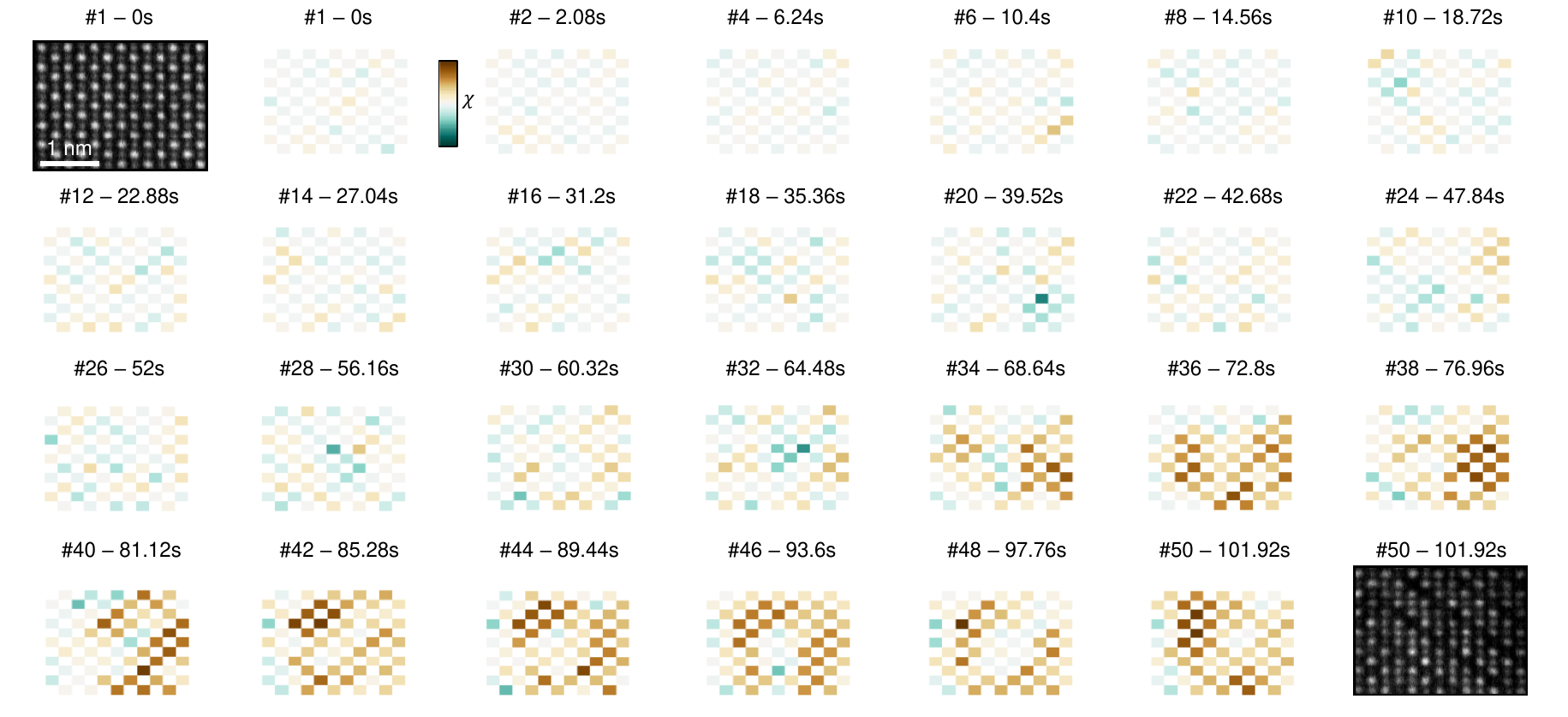}}
\renewcommand{\figurename}{Extended Data Fig.|}
\caption{\label{SIfigcorr}
\textbf{Time- and position-dependent initiation of the structural transformation.}
The degree of ordering is calculated from nearest neighbor Cr atom column intensities (see Extended Data Fig.~\ref{SIfig6}). The differences along diagonal bottom left to top right are defined as $\Delta_{15} = |I_1 - I_5|$ and $\Delta_{25} = |I_2 - I_5|$ while along diagonal top left to bottom right are defined as $\Delta_{35} = |I_3 - I_5|$ and $\Delta_{45} = |I_4 - I_5|$. A degree of ordering is then defined as $\chi = (\Delta_{15} + \Delta_{35}) - (\Delta_{25} + \Delta_{45})$.
}
\end{figure*}
%##############################################################################
%
\newpage
%
%###################### Figure 1 ################################################
\begin{figure*}[ht]
\scalebox{\figurescale}{\includegraphics[width=1\linewidth]{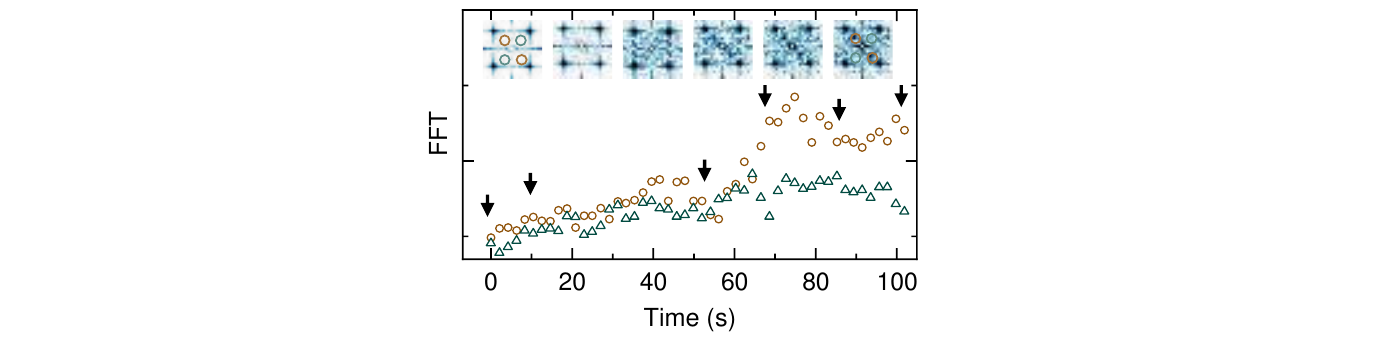}}
\renewcommand{\figurename}{Extended Data Fig.|}
\caption{\label{SIfigFFT}
\textbf{Time evolution of Cr rearrangement in Fourier space.} 
The intensity of the double periodicity spots in the FFT from a real-space image is integrated as a function of time. The spots are an indication of Cr rearrangement along the two different diagonals. Inset: Corresponding FFT images at $\SI{0}{\second}$, $\SI{10.40}{\second}$, $\SI{56.16}{\second}$, $\SI{68.64}{\second}$, $\SI{81.12}{\second}$ and $\SI{101.92}{\second}$. Similar to the real space analysis, there is a continuous increase for both orientations that changes into a preferred rearrangement along one Cr diagonal at $\SI{62}{\second}$.
}
\end{figure*}
%##############################################################################
%
\newpage

%
%###################### Figure 7 ################################################
%\begin{figure}[ht]
%\scalebox{\figurescale}{\includegraphics[width=1\linewidth]{Overview_lines_stacking_faults.pdf}}
%\renewcommand{\figurename}{Extended Data Fig.|}
%\caption{\label{SIfig7}
%
%\textbf{a}, Low-magnification STEM-HAADF image of an electron beam dosed area reveals two different types of defect structures highlighted by the colored squares. Inset: Corresponding FFT of the STEM-HAADF image.
%\textbf{b}, Type A: Maze like defect structure of 1D lines along the [110] direction forming a superlattice structure that shows up as a double periodicity peak in the FFT.
%\textbf{c}, Type B: Stacking fault defect structure that is accompanied by reduced STEM contrast attributed to the loss of material, very likely Br atoms. The crystallography of this defect structure is discussed in greater detail in SI Fig.~\ref{SIfig8}.
%}
%\end{figure}
%##############################################################################
%

%
%###################### Figure 9 ################################################
\begin{figure*}[ht]
\scalebox{\figurescale}{\includegraphics[width=1\linewidth]{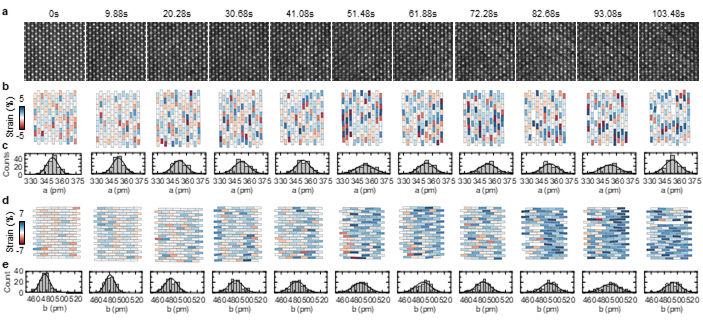}}
\renewcommand{\figurename}{Extended Data Fig.|}
\caption{\label{SIfig9}
\textbf{Time evolution of the 2D superlattice distortion kinetics and resulting net strain.} 
\textbf{a}, STEM-HAADF images of multilayer CrSBr taken at $\SI{200}{\kilo\electronvolt}$.
\textbf{b}, Strain maps along the $a$-direction taking into account the positions of the S/Br atom columns.
\textbf{c}, Corresponding $a$-distance histogram.
\textbf{d}, Strain maps along the $b$-direction taking into account the positions of the S/Br atom columns.
\textbf{e}, Corresponding $b$-distance histogram.
}
\end{figure*}
%##############################################################################
%
\newpage

\section{Twinning defects.}

%
%###################### Figure 9 ################################################
\begin{figure*}[ht]
\scalebox{\figurescale}{\includegraphics[width=1\linewidth]{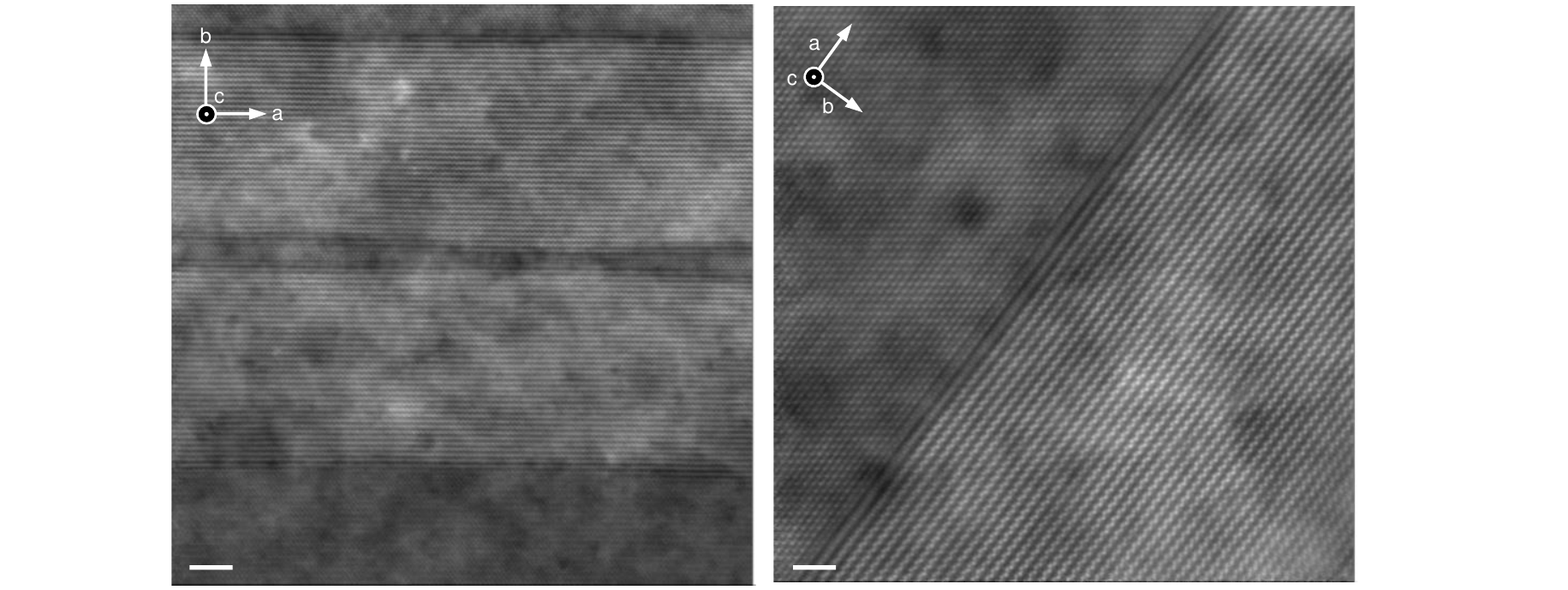}}
\renewcommand{\figurename}{Extended Data Fig.|}
\caption{\label{SIfiggrain}
\textbf{Twinning in CrSBr.} 
STEM-HAADF images of grain boundaries in CrSBr recorded at $\SI{60}{\kilo\electronvolt}$.
}
\end{figure*}
%##############################################################################
%

\newpage

\section{Stacking fault defect.}

%
%###################### Figure 8 ################################################
\begin{figure*}[ht]
\scalebox{\figurescale}{\includegraphics[width=1\linewidth]{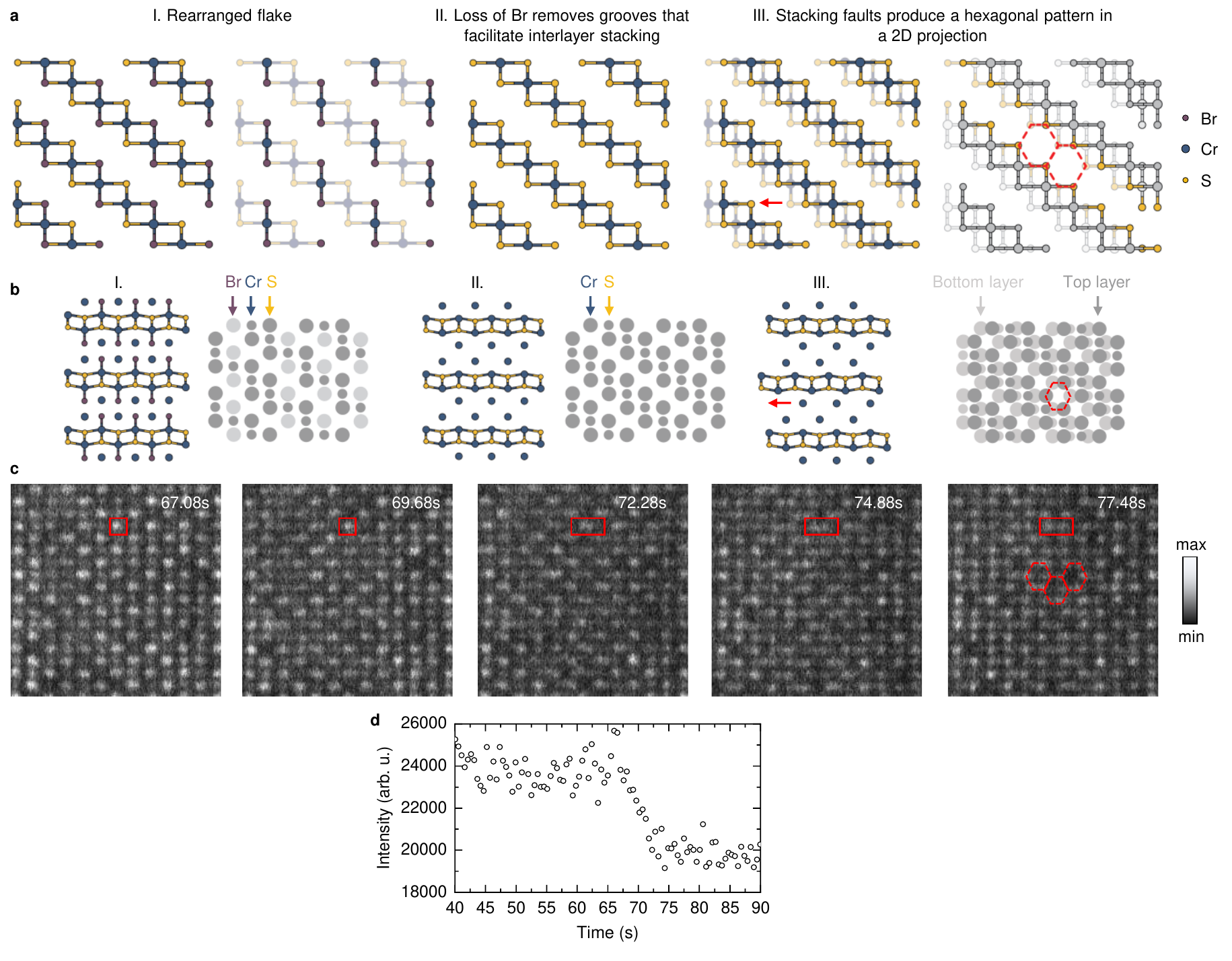}}
\renewcommand{\figurename}{Extended Data Fig.|}
\caption{\label{SIfig8}
\textbf{Type 2 irreversible stacking fault defect structure kinetics.} 
\textbf{a}, and \textbf{b}, Top view and side view schematic illustration of one possible crystallographic arrangement of atoms and individual atom planes under electron beam irradiation. Faded colors indicate multiple layers in which the bottom layers are shown with lower intensity. I. Type 1 rearranged structure. II. Beyond the regime where the type 1 rearranged structure forms, extended electron irradiation can result in the loss of atoms, presumably Br atoms forming Br$_2$ gas from proximal Br atoms which is released to the vacuum. III. Potential loss of Br atoms results in an increased mobility of individual layers consisting mostly of S and Cr that shift with respect to one another along the $a$-direction by $~\SI{1.6}{\angstrom}$ showing a distorted hexagonal lattice pattern as a consequence.
\textbf{c}, STEM-HAADF images for selected times taken from a sequence of data acquired while the type 2 defective structure is formed. A S/Br column is highlighted that appears to split in two columns as time elapses. The hexagonal pattern is highlighted by the red hexagons.
\textbf{d}, Integrated atom column intensity as a function of time suggests knock-on damage and removal of material simultaneous with the formation of the type 2 defect structure.
}
\end{figure*}
%##############################################################################
%

\newpage

\section{Cr migration and self-healing.}

%
%###################### Figure 9 ################################################
\begin{figure*}[ht]
\scalebox{\figurescale}{\includegraphics[width=1\linewidth]{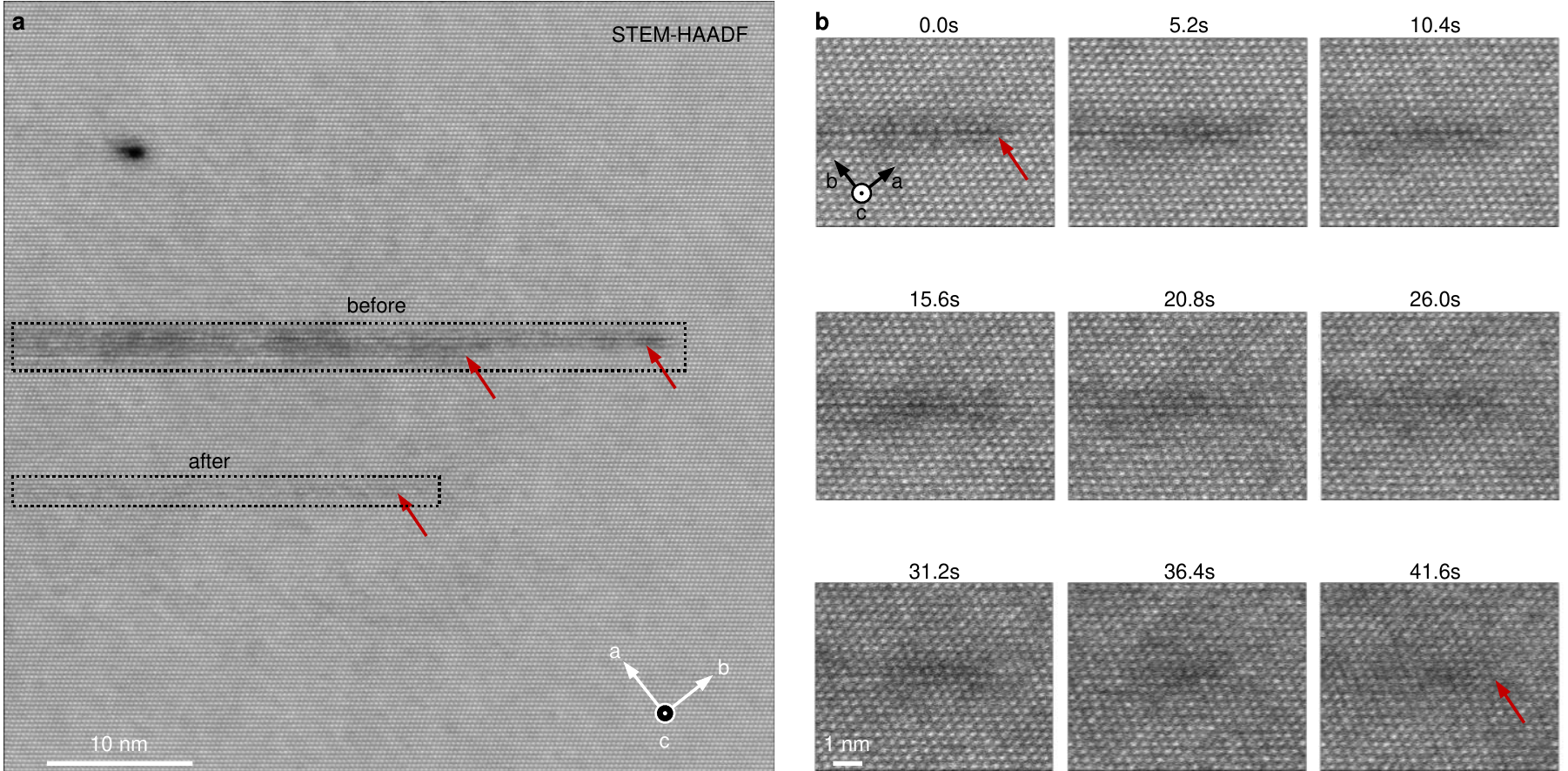}}
\renewcommand{\figurename}{Extended Data Fig.|}
\caption{\label{SIfigcut_video}
\textbf{Self-healing of electron beam induced Cr vacancy line in CrSBr.} 
\textbf{a}, STEM-HAADF images Cr vacancy lines created by scanning the electron probe along the Cr diagonal. Two lines on the top are imaged before subsequent imaging and appear as dark lines. The line on the bottom was imaged for $~\SI{40}{\second}$ after generating the Cr vacancy line. The image is taken with an electron beam energy of $\SI{200}{\kilo\electronvolt}$. The Cr atoms migrate into the vdW gap due to the energy provided by the electrons resulting in a self-healing of the vacancy line when imaged.
\textbf{b}, STEM-HAADF image series for an electron beam energy of $\SI{200}{\kilo\electronvolt}$ after generating a 1D Cr vacancy line.
}
\end{figure*}
%##############################################################################
%

\newpage

\section{CrSBr material characterization.}

%
%###################### Figure 10 ################################################
\begin{figure*}[ht]
\scalebox{\figurescale}{\includegraphics[width=1\linewidth]{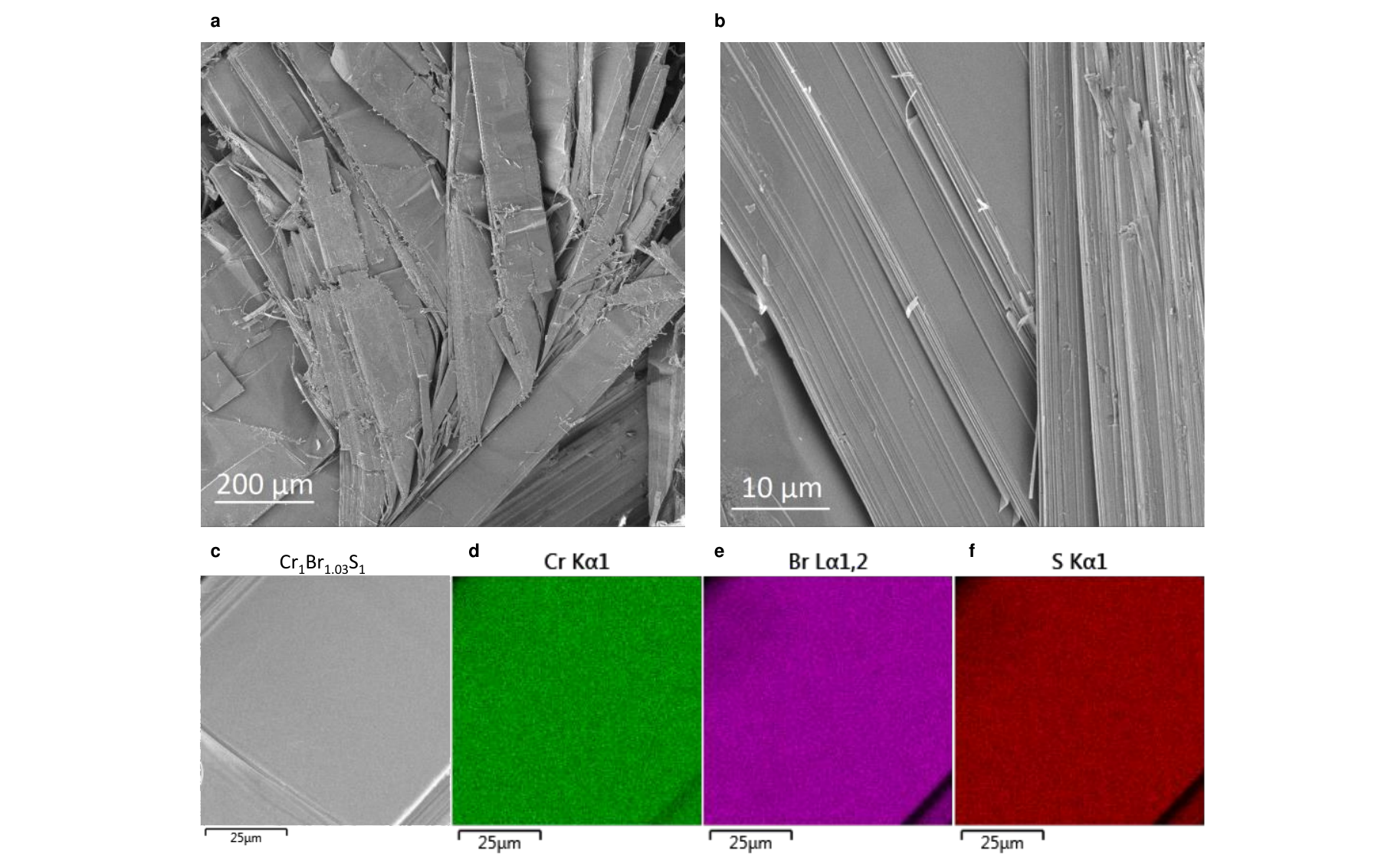}}
\renewcommand{\figurename}{Extended Data Fig.|}
\caption{\label{SIfigEDS}
\textbf{SEM and EDS of bulk CrSBr} 
\textbf{a} and \textbf{b}, SEM image of bulk CrSBr showing the layered nature and crystal anisotropy. The SEM images shows platelet character typical for van der Waals layerd materials. The EDS shows homogeneous distribution of Cr, Br and S. 
\textbf{c}, SEM image of bulk CrSBr.
\textbf{d-f}, Corresponding EDS maps of the same region.
}
\end{figure*}
%##############################################################################
%
\newpage

%
%###################### Figure 11 ################################################
\begin{figure*}[ht]
\scalebox{\figurescale}{\includegraphics[width=1\linewidth]{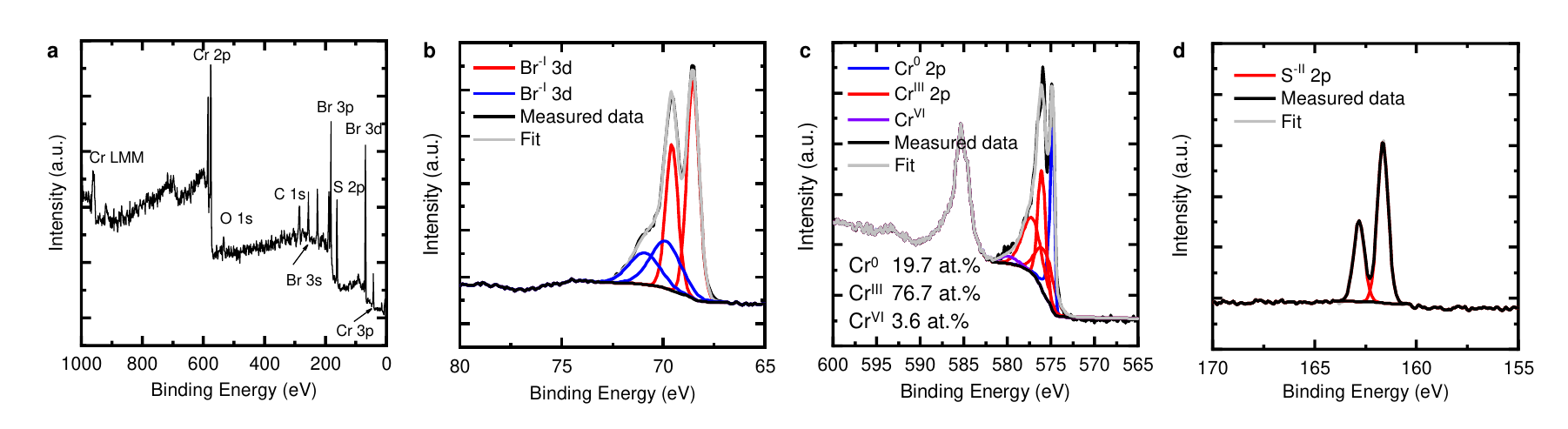}}
\renewcommand{\figurename}{Extended Data Fig.|}
\caption{\label{SIfigXPS}
\textbf{XPS of bulk CrSBr} 
\textbf{a-d}, High resolution XPS shows presence of Br$^-$ and S$^{2-}$ anions and chromium primarily in the $3+$ oxidation state. 
}
\end{figure*}
%##############################################################################
%

\newpage
%
%###################### Figure 12 ################################################
\begin{figure*}[ht]
\scalebox{\figurescale}{\includegraphics[width=1\linewidth]{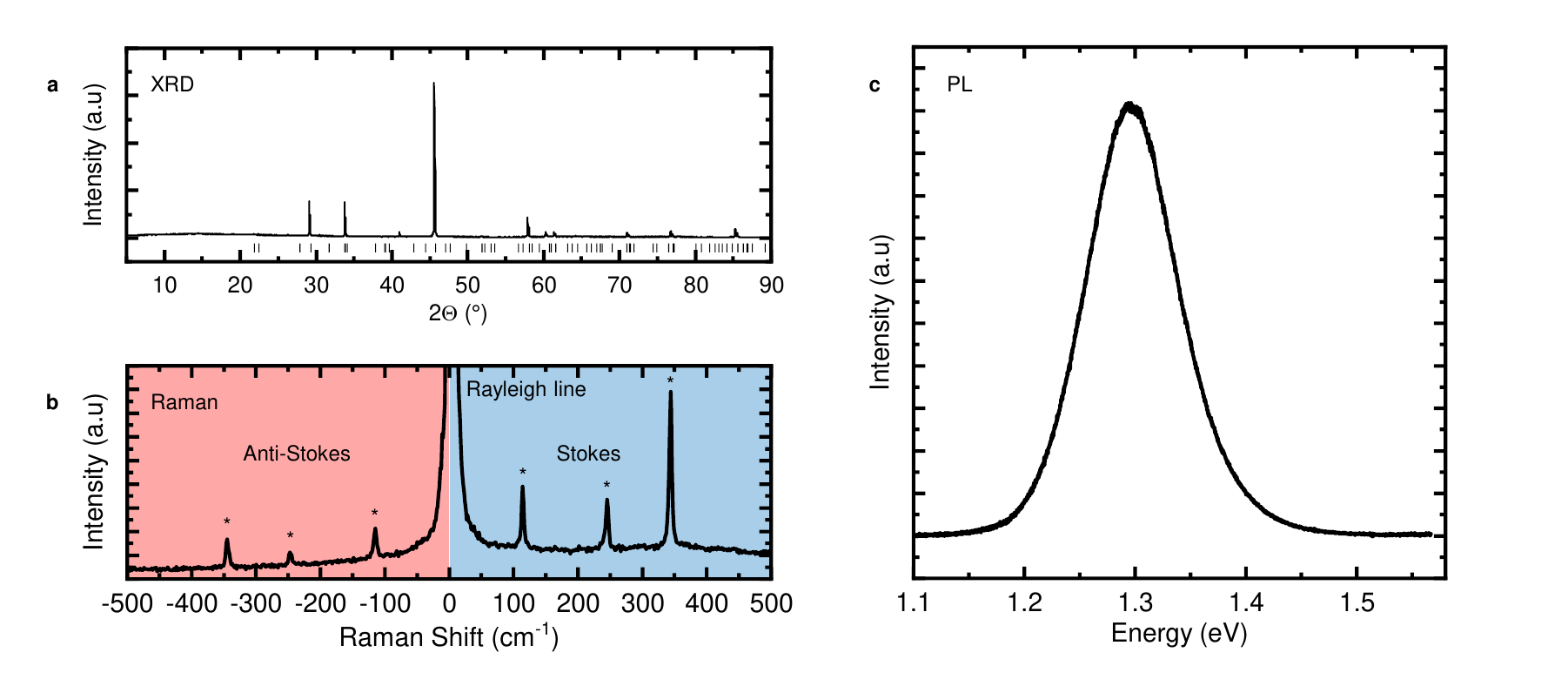}}
\renewcommand{\figurename}{Extended Data Fig.|}
\caption{\label{SIfigRamanPLXRD}
\textbf{XRD, Raman and PL of bulk CrSBr} 
\textbf{a}, X-ray diffraction data corresponds to the pure single phase CrSBr with high preferential orientation due to layered structure.
\textbf{b}, Room temperature Raman spectrum of multilayer CrSBr recorded with an excitation wavelength of $\SI{532}{\nano\meter}$ and a low-frequency filter set.
\textbf{c}, Room temperature of multilayer CrSBr recorded with an excitation wavelength of $\SI{785}{\nano\meter}$.
}
\end{figure*}
%##############################################################################
%

\newpage

%
%##############################################################################
%               Acknowledgements & Contributions
%##############################################################################
%
%###############################################################################
%								Additional information
%###############################################################################
%
%###############################################################################
%								BIBLIOGRAPHY
%##############################################################################
%
%\FloatBarrier
\bibliographystyle{apsrev}
\bibliography{full}% Produces the bibliography via BibTeX.